\documentclass[11pt,a4paper]{article}

\usepackage[margin=2.2cm]{geometry}
\usepackage[T1]{fontenc}
\usepackage{lmodern}
\usepackage{amsmath,amssymb}
\usepackage{graphicx}
\usepackage{xcolor}
\usepackage{booktabs,multirow,tabularx,float,threeparttable}
\usepackage[normalem]{ulem}
\usepackage{xurl}
\usepackage{etoolbox}
\usepackage{authblk}
\usepackage[numbers,square,sort&compress]{natbib}
\usepackage{hyperref}
\hypersetup{colorlinks=true,linkcolor=blue,citecolor=blue,urlcolor=blue}
\graphicspath{{./}{./figures/}}
\usepackage{dblfloatfix}

\usepackage{tikz}
\usetikzlibrary{arrows,shapes,intersections,positioning,shadows,arrows.meta,fit,backgrounds,calc,shapes.geometric}
\usepackage{placeins}
\usepackage{makecell}
\usepackage[acronym,nopostdot,nonumberlist,hyperfirst=false]{glossaries}
\usepackage{accsupp}

\newenvironment{acks}{\section*{Acknowledgements}}{}
\newenvironment{ethics}{\section*{Ethical Considerations}}{}
\newenvironment{openscience}{\section*{Data Availability Statement}}{}

\usepackage[textsize=scriptsize]{todonotes}

\newcommand{\mm}[1]{\todo[inline,color=red]{MM}}

\definecolor{ForestGreen}{RGB}{34,139,34}

\newcommand*\halfcirc[1][0.8ex]{%
  \BeginAccSupp{method=hex,unicode,ActualText=25D1}%
  \begin{tikzpicture}
    \draw[fill] (0,0)-- (90:#1) arc (90:270:#1) -- cycle ;
    \draw[thick] (0,0) circle (#1);
  \end{tikzpicture}%
  \EndAccSupp{}%
}

\newcommand*\fullcirc[1][0.8ex]{%
  \BeginAccSupp{method=hex,unicode,ActualText=25CF}%
  \tikz\fill (0,0) circle (#1);%
  \EndAccSupp{}%
}

\makeglossaries
\newacronym{ieee}{IEEE}{Institute of Electrical and Electronics Engineers}
\newacronym{acm}{ACM}{Association for Computing Machinery}
\newacronym{w3c}{W3C}{World Wide Web Consortium}
\newacronym{ukicc}{UK ICC}{United Kingdom International Chamber of Commerce}
\newacronym{ico}{ICO}{Information Commissioner's Office}
\newacronym{rnib}{RNIB}{Royal National Institute of Blind People}
\newacronym{webaim}{WebAIM}{Web Accessibility in Mind}

\newacronym{gdpr}{GDPR}{General Data Protection Regulation}
\newacronym{ukgdpr}{UK GDPR}{UK General Data Protection Regulation}
\newacronym{dpa}{DPA 2018}{Data Protection Act 2018}
\newacronym{pecr}{PECR}{Privacy and Electronic Communications Regulations}
\newacronym{duaa}{DUAA}{Data Use and Access Act 2025}
\newacronym{ccpa}{CCPA}{California Consumer Privacy Act}
\newacronym{cpra}{CPRA}{California Privacy Rights Act}
\newacronym{pipl}{PIPL}{Personal Information Protection Law}
\newacronym{lgpd}{LGPD}{Lei Geral de Proteção de Dados}
\newacronym{pipeda}{PIPEDA}{Personal Information Protection and Electronic Documents Act}
\newacronym{epd}{ePD}{ePrivacy Directive}
\newacronym{eaa}{EAA}{European Accessibility Act (Directive (EU) 2019/882)}

\newacronym{blv}{BLV}{Blind and Low Vision}
\newacronym[plural={ATs}, longplural={Assistive Technologies}]{at}{AT}{Assistive Technology}
\newacronym[
  short={VAT},
  shortplural={VATs},
  long={Visual \acrlong{at}},
  longplural={Visual \acrlongpl{at}}
]{vat}{VAT}{Visual Assistive Technology}
\newacronym[plural={VIs}, longplural={Visual Impairments}]{vi}{VI}{Visual Impairment}
\newacronym{ai}{AI}{Artificial Intelligence}
\newacronym{llm}{LLM}{large language model}

\newacronym{uk}{UK}{United Kingdom}
\newacronym{eu}{EU}{European Union}

\newacronym{nvda}{NVDA}{Non-Visual Desktop Access}
\newacronym{jaws}{JAWS}{Job Access With Speech}

\newacronym{pii}{PII}{personally identifiable information}

\newacronym{webgl}{WebGL}{Web Graphics Library}
\newacronym[plural={APIs},longplural={Application Programming Interfaces}]{api}{API}{Application Programming Interface}
\newacronym{js}{JS}{JavaScript}
\newacronym{html}{HTML}{HyperText Markup Language}
\newacronym{dom}{DOM}{Document Object Model}
\newacronym{css}{CSS}{Cascading Style Sheets}
\newacronym[plural={URLs}, longplural={Uniform Resource Locators}]{url}{URL}{Uniform Resource Locator}
\newacronym{webrtc}{WebRTC}{Web Real-Time Communication}
\newacronym{http}{HTTP}{Hypertext Transfer Protocol}
\newacronym{https}{HTTPS}{Hypertext Transfer Protocol Secure}
\newacronym{dns}{DNS}{Domain Name System}
\newacronym{cdn}{CDN}{Content Delivery Network}
\newacronym[plural={IPs}, longplural={Internet Protocol addresses}]{ip}{IP}{Internet Protocol}
\newacronym[plural={CMPs}, longplural={Consent management platforms}]{cmp}{CMP}{Consent management platform}
\newacronym{captcha}{CAPTCHA}{Completely Automated Public Turing test to tell Computers and Humans Apart}
\newacronym{ssl}{SSL}{Secure Sockets Layer}
\newacronym{ui}{UI}{User Interface}

\newacronym{cws}{CWS}{Chrome Web Store}
\newacronym{sp}{SP}{Security and Privacy}
\newacronym{snes}{SNEs}{Security-Noteworthy Extensions}
\newacronym{wave}{WAVE}{WAVE Web Accessibility Evaluation Tool}
\newacronym{wcag}{WCAG}{Web Content Accessibility Guidelines}
\newacronym{aria}{ARIA}{Accessible Rich Internet Applications}
\newacronym{cms}{CMS}{Content Management Systems}
\newacronym{vv8}{VV8}{VisibleV8}
\newacronym{csp}{CSP}{Content Security Policy}
\newacronym[plural={VPNs}, longplural={Virtual Private Networks}]{vpn}{VPN}{Virtual Private Network}

\newacronym{cv}{CV}{Coefficient of Variation}

\let\glsgenacfmtOrig\glsgenacfmt
\renewcommand*{\glsgenacfmt}{%
  \ifglsused\glslabel
    {\glsgenacfmtOrig}%
    {\hypertarget{glo:\glslabel}{\glsgenacfmtOrig}}%
}

\renewcommand*{\glsdohyperlink}[2]{%
  \glsdoshowtarget{#1}%
  {%
    \glsdohyperlinkhook{#1}{#2}%
    \begingroup
      \hypersetup{linkcolor=black}%
      \hyperlink{#1}{#2}%
    \endgroup
  }%
}
\makeatletter
\let\@glslink\glsdohyperlink
\makeatother

\begin{document}

\title{A Large-Scale Evaluation of Cookie Notice Privacy and Accessibility for Blind and Low Vision Users}

\author[1]{James M. Clarke}
\author[2]{Maryam Mehrnezhad}
\affil[1]{Surrey Centre for Cyber Security, University of Surrey, Guildford, United Kingdom}
\affil[2]{Information Security Group (ISG), Royal Holloway University of London, Egham, United Kingdom}
\affil[ ]{Corresponding author: Maryam Mehrnezhad (\href{mailto:Maryam.Mehrnezhad@rhul.ac.uk}{Maryam.Mehrnezhad@rhul.ac.uk})}
\date{}

\maketitle
\begin{abstract}
We present the first large-scale privacy and accessibility evaluation of cookie notices across 2,061 domains for blind and low vision (BLV) users.
Using a Playwright-based crawler, a custom NVDA screen reader add-on, and the Lighthouse and WAVE accessibility tools, we measure UK GDPR/PECR compliance, tracker activity across pre-interaction, post-accept, and post-reject phases, and cookie notice accessibility for screen reader users. We find that 51\,\% of sites display a cookie notice, of which (at least) 50\,\% violate UK GDPR/PECR requirements.
Accepting a notice produces a mean five-fold increase in tracker activity, and degrades page landmark structure on one in five sites. While 95\,\% of notices are readable by a screen reader, only 37\,\% carry a meaningful title, 39\,\% are announced immediately on page load, and 42\,\% use descriptive button labels. We reveal a systemic correlation: cookie notices failing legal privacy requirements disproportionately exhibit the weakest screen reader accessibility. This compounding privacy burden on BLV users demands urgent intervention from developers, regulators, and standards bodies.
\end{abstract}

\section{Introduction}
Cookie notices are the primary mechanism through which websites obtain user consent under data protection law~\cite{matte2020cookie,GeneralDataProtection,mehrnezhadCrossPlatformEvaluationPrivacy2020}. For \gls{blv} users, they present a particular challenge. Unlike sighted users, screen reader users cannot visually scan or locate a notice and must instead navigate to it sequentially~\cite{gerber_surfing_2002}. This means that \gls{blv} users may accept or miss a cookie notice potentially without being fully aware, leading to further privacy implications.

To the best of our knowledge, no prior work combines large-scale \gls{ukgdpr} and \gls{pecr} compliance measurement, post-interaction tracking analysis, and \gls{blv} accessibility evaluation within a single study. Existing automated compliance audits do not consider screen reader accessibility, and existing accessibility studies do not examine how consent interactions affect a page's network behaviour or are small-scale and manual ~\cite{bouhoulaAutomatedLargeScale2023,nouwensDarkPatternsGDPR2020,matte2020cookie,kretschmerCookieBannersPrivacy2021,trevisanYearsEUCookie2019, ClarkeInvisible2024}. The relationship between a notice's privacy compliance status and its accessibility has not been studied. It is also unclear what happens to page accessibility after a cookie notice is dismissed, including whether landmark structure is preserved across interaction states.

Our research questions include: 
\begin{itemize}
  \item
\textbf{RQ1}: How do cookie notices on surveyed websites meet \textbf{\gls{ukgdpr} and \gls{pecr} requirements}? {RQ1.1}: How do websites alter their tracking behaviour across consent phases (pre-interaction, post-accept, post-reject)? {RQ1.2}: 
What are the common response options of cookie notices, and do they align with those required by \gls{ukgdpr} and \gls{pecr}? 
\item
\textbf{RQ2}: To what extent do cookie notices meet \textbf{\gls{wcag} 2.2 accessibility requirements} for screen reader users? {RQ2.1}: How do cookie notices affect the automated \gls{wcag} 2.2 accessibility scores of websites? {RQ2.2}: To what extent are cookie notices accessible with screen readers? 
\item
\textbf{RQ3}: What is the \textbf{relationship} between cookie notice accessibility and compliance with \gls{ukgdpr} and \gls{pecr}? {RQ3.1}:  Are cookie notices that are compliant more likely to be accessible to screen reader users?
{RQ3.2}: Do cookie notices that are compliant have higher automated accessibility scores than non-compliant notices?
\end{itemize}

\begin{figure}[t]
  \centering
  \includegraphics[width=0.8\linewidth,keepaspectratio]{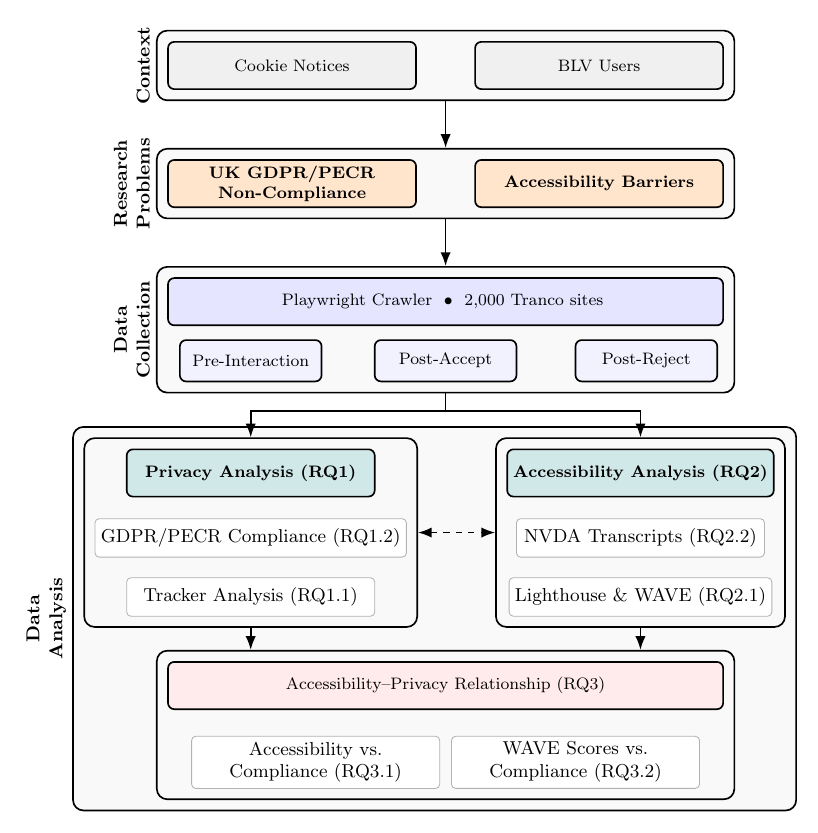}
  \caption{Conceptual diagram for automated analysis of the accessibility and privacy of cookie notices.}
  \label{fig:cookie_framework}
\end{figure}

\autoref{fig:cookie_framework} provides a conceptual overview of our approach, illustrating how we integrate automated crawling, screen reader analysis, and accessibility to address our research questions.

Our contributions are: 

\textbf{(1)} A large-scale automated crawler covering 3,000 sites with three interaction phases (pre-interaction, post-accept, post-reject), showing that 49\% of the 2,061 reachable sites carry no cookie notice (a violation), and that accepting a notice produces a mean five-fold increase in tracker activity. 

\textbf{(2)} A custom \gls{nvda} add-on for automated screen reader transcript capture, enabling analysis of how cookie notices are read and navigated by \gls{at}; we find that while 95\,\% of notices are readable, only 37\,\% carry a meaningful title, 39\,\% are announced immediately on page load, and 42\,\% use descriptive button labels. 

\textbf{(3)}
An empirical compliance and accessibility audit of over 2,000 reachable websites (identifying 1,058 cookie notices), covering \gls{ukgdpr} and \gls{pecr} compliance rates, tracking behaviour changes across consent phases, and the accessibility quality of cookie notices for \gls{blv} users at scale, finding that 50\,\% of sites with a detectable notice violate at least one key legal requirement: 33\,\% offer no direct reject path and a further 17\,\% offer a reject path but visually de-emphasise it. 

\textbf{(4)} A differential accessibility analysis method that compares automated accessibility scores across interaction phases to identify cookie notice-related accessibility issues referencing \gls{wcag} 2.2, revealing that accepting a notice degrades page landmark structure in one in five sites.

\begin{table*}[t]\centering\footnotesize
    \setlength{\tabcolsep}{1.5pt}
    \caption{Overview of example work investigating cookie notices and/or accessibility of websites.}
    \label{tab:cookie_related_work}
    \footnotesize
    \begin{threeparttable}
    \begin{tabular}{>{\fontsize{7}{9}\selectfont}l*{9}{>{\fontsize{7}{9}\selectfont}c}}
        \toprule
        \textbf{Work} & \textbf{Year} & \makecell{\textbf{No. of} \\ \textbf{Sites}} &
       \makecell{ \textbf{Auto-} \\ \textbf{mation}}
       & \makecell{\textbf{Privacy} \\ \textbf{Compliance}} & \makecell{\textbf{Tracker} \\ \textbf{Analysis}} & \makecell{\textbf{Dark} \\ \textbf{Patterns}} &
        \makecell{\textbf{Post-Interaction} \\ \textbf{Analysis}} &
        \makecell{\textbf{Accessibility}\\ \textbf{Scoring}} & \makecell{\textbf{Screen Reader} \\ \textbf{Analysis}} \\ \midrule
        Trevisan et al.~\cite{trevisanYearsEUCookie2019}     & 2019 & 35{,}000  & \fullcirc & \fullcirc & \fullcirc &           & \halfcirc &           & \\
        Utz et al.~\cite{utzInformedConsentStudying2019}      & 2019 & 1{,}000   & \halfcirc & \fullcirc &           & \halfcirc &           &           & \\
        Degeling et al.~\cite{degelingWeValueYour2019}        & 2019 & 6{,}759   & \halfcirc & \halfcirc & \halfcirc &           &           &           & \\
        Nouwens et al.~\cite{nouwensDarkPatternsGDPR2020}     & 2020 & 10{,}000  & \halfcirc & \fullcirc &           & \fullcirc &           &           & \\
        Matte et al.~\cite{matte2020cookie}                   & 2020 & 28{,}257  & \fullcirc & \fullcirc & \fullcirc & \halfcirc & \fullcirc &           & \\
        Schnell and Roy~\cite{schnellWebsitePrivacyNotification2021} & 2021 & 40 & \halfcirc & \halfcirc &           &           &           & \halfcirc & \halfcirc \\
        Kampanos et al.~\cite{kampanos2021accept}   & 2021 & 17{,}000  & \halfcirc & \fullcirc & \halfcirc & \halfcirc &           &           & \\
        Bollinger et al.~\cite{bollinger2022automating}        & 2022 & 30{,}000  & \fullcirc & \fullcirc & \fullcirc & \halfcirc & \fullcirc &           & \\
        Rasaii et al.~\cite{rasaii2023exploring}              & 2023 & 10{,}000  & \fullcirc & \fullcirc & \fullcirc & \halfcirc & \fullcirc &           & \\
        Bouhoula et al.~\cite{bouhoulaAutomatedLargeScale2023} & 2023 & 97{,}000 & \fullcirc & \fullcirc & \fullcirc & \halfcirc & \fullcirc &           & \\
        Clarke et al.~\cite{ClarkeInvisible2024} & 2024 & 46        & \halfcirc & \halfcirc &           & \halfcirc &           & \fullcirc & \fullcirc \\
        \midrule
        This work                                             & 2026 & 3{,}000   & \fullcirc & \fullcirc & \fullcirc & \fullcirc & \fullcirc & \fullcirc & \fullcirc \\
        \bottomrule
    \end{tabular}
    \begin{tablenotes}\footnotesize
      \item \fullcirc~= Full analysis, \halfcirc~= Partial analysis.
    \end{tablenotes}
    \end{threeparttable}
\end{table*}

\section{Background and related work} \label{sec:background}
This section provides the  UK/EU legal framework governing cookie consent, the \gls{blv} population and the AT landscape, and prior work on these topics.

\subsection{Cookie notices and privacy law} \label{sec:laws}
The \gls{gdpr}, implemented in May 2018, introduced significant changes to the rules governing online tracking and consent, including consent to cookies~\cite{GeneralDataProtection,matte2020cookie}.
To be valid, consent must be freely given, specific, informed, unambiguous, explicit, revocable, obtained prior to any data collection, and requested in a readable and accessible manner~\cite{GeneralDataProtection,gerl2019layered}.
The ePrivacy Directive (``\gls{epd}'', commonly referred to as the ``cookie law'')~\cite{Directive2002582002} provides supplementary rules to the \gls{gdpr}. Under the \gls{epd}, publishers must obtain user consent before collecting and processing personal data via non-mandatory cookies or similar technologies, that is, those not strictly necessary for the services requested by the user~\cite{Directive2002582002}. There is ongoing legislative activity to replace the \gls{epd} with the ePrivacy Regulation, which will harmonise privacy rules across the EU and introduce stricter requirements for consent and data handling~\cite{ePrivacyReg2017}.

The \gls{ukgdpr}, retained in UK law via the European Union (Withdrawal) Act 2018, defines the standard for valid consent and the principles of lawful data processing in the UK~\cite{UKGDPR}. The \gls{dpa} supplements the \gls{ukgdpr}, providing the domestic legislative framework for data protection, including data subject rights and enforcement mechanisms~\cite{DataProtectionAct2018}. The Privacy and Electronic Communications Regulations (\gls{pecr}) govern the use of cookies and similar technologies in the UK, requiring that users be informed about the purposes of data collection and that consent be obtained for non-essential cookies~\cite{PECR}. The consent standard under \gls{pecr} must meet the \gls{ukgdpr} threshold.
The \gls{duaa}, the majority of whose cookie-related provisions came into force in Feb 2026, amended \gls{pecr} to introduce new exemptions for certain low-risk cookie uses, specifically analytics cookies used to improve a service, and cookies used for security and fraud detection~\cite{DataUseAccessAct2025}. For these exempt categories, operators must still provide clear information and offer users the right to opt out. 

Prior to the UK's departure from the EU, UK cookie law was governed by the same framework: the \gls{epd} implemented domestically via \gls{pecr}, with consent standards derived from the \gls{gdpr}. Following Brexit, the UK retained this framework through the European Union (Withdrawal) Act 2018, creating parallel but formally independent instruments: the \gls{ukgdpr} and the amended \gls{pecr}. At the point of departure, the two frameworks were substantively identical; they have since begun to diverge. In the EU, the \gls{epd} remains in force and consent is required for all non-essential cookies; efforts to replace it with the ePrivacy Regulation remain ongoing and unadopted~\cite{ePrivacyReg2017}. In the UK, the \gls{duaa} amended \gls{pecr} to introduce new exemptions from consent for certain low-risk purposes, including analytics and customisation, which have no equivalent in the current EU framework~\cite{DataUseAccessAct2025}. 

The practical consequence for organisations operating across both jurisdictions is that compliance with the stricter EU standard does not guarantee UK compliance, and vice versa, as the two regimes now impose distinct obligations. This divergence is particularly relevant to this paper, with the crawls conducted from the UK but with implications for the broader EU context. 

\subsection{Cookie notice dark patterns}
A body of research exists on the implementation and effectiveness of cookie notices, privacy banners, and tracking practices~\cite{utzInformedConsentStudying2019,degelingWeValueYour2019,mehrnezhadCrossPlatformEvaluationPrivacy2020,trevisanYearsEUCookie2019,mehrnezhadHowCanWould2022,huCharacterisingThirdParty2019,englehardtOnlineTracking1millionsite2016,sorensenGDPRChangesThird2019,GDPRwhatHappened,rasaii2023cookiewall,rasaii2023exploring,zhang_cschecker_2024,gray_dark_2018}. Identified dark patterns include: providing invalid consent, nudging users towards acceptance, making opt-out difficult or unavailable, failing to allow users to withdraw previously given consent, pre-enabling non-essential cookies, and embedding trackers within the cookie notice itself. The top five consent management platforms have been found to employ such dark patterns~\cite{nouwensDarkPatternsGDPR2020}. 

Tracking has been shown to persist even when users reject cookie notices~\cite{ormeno_cookie_2024}.
Research has also examined the user dimensions of tracking, including negative attitudes towards tracking practices~\cite{rao2015they,coopamootoo2017privacy}, demographic differences such as gender and country~\cite{coopamootooFeelInvadedAnnoyed}, and the gap between regulation, real-world website practices, and users' limited ability to protect themselves, alongside their demand for greater transparency and control~\cite{mehrnezhadHowCanWould2022,shiraziWhatDetersJane2014,pugliese2020long,melicherPreferencesWebTracking2016,urSmartUsefulScary2012}.
Furthermore, nudging has been shown to be highly effective in influencing privacy disclosure decisions~\cite{ioannou2021privacy,bermejo2021website}.

An example of large-scale work is~\cite{bouhoulaAutomatedLargeScale2023}, where 97k EU websites were studied, using NLP to interact with cookie notices and compare declared purposes to actual cookie usage, finding that 65.4\% of websites with a reject option still place analytics or advertising cookies after explicit negative consent, and that 90.2 \% violate at least one \gls{gdpr} requirement~\cite{bouhoulaAutomatedLargeScale2023}. This study, however, does not address screen reader accessibility for \gls{blv} users, and it predates the UK's \gls{duaa} amendments to \gls{pecr}; this paper addresses both gaps. \autoref{tab:cookie_related_work} summarises some of these works and situates this paper as the first large-scale automated study of 2,061 websites for privacy compliance, accessibility metrics, and their correlations.

\subsection{Visual impairments and AT standards}
There are an estimated 285 million people worldwide with a \gls{vi}~\cite{dosono_toward_2018}, and approximately 20\,\% of the global population relies on \gls{at}~\cite{palani_web_2022}. The UK's \gls{rnib} reports over two million people living with sight loss~\cite{rnibKeyStats}, around 19\,\% of whom use screen readers ~\cite{rnibViLives2022}. Among desktop screen readers, \gls{nvda} is the most commonly used overall (65.6\,\% vs 60.5\,\% for \gls{jaws}), with 71.6\,\% of users having more than one screen reader tool~\cite{ScreenReaderUserSurvey}. \gls{blv} users' reliance on \glspl{at} creates a distinct set of privacy and accessibility challenges on the web.
Fundamentally different from sighted users, \gls{blv} users primarily employ either a \emph{searching} strategy (heading or link navigation) or a \emph{scrolling} strategy (listening to page content sequentially)~\cite{gerber_surfing_2002}, and real-world user behaviour on web applications confirms a strong dependence on landmark-based scanning over logical sequential navigation~\cite{takagi_analysis_2007}. A direct implication is that a notice that is not the first thing for the screen reader, or that lacks a structural heading to identify the consent region, forces \gls{blv} users to either scan the entire page or miss the notice entirely.

The W3C \gls{wcag} is the internationally recognised standard for web accessibility~\cite{w3WCAGOverview2022}, organising requirements under four principles (perceivable, operable, understandable, and robust) at three conformance levels: A (minimum), AA (standard), and AAA (highest). In the UK, the Equality Act 2010 imposes a general duty on service providers not to discriminate against disabled people and to make reasonable adjustments, without prescribing a specific technical conformance level~\cite{equalityact2010}, while the Public Sector Bodies (Websites and Mobile Applications) (No.~2) Accessibility Regulations 2018 require public-sector websites to meet \gls{wcag} 2.1 Level AA~\cite{psbar2018,en301549}. At the EU level, the \gls{eaa} extends comparable requirements to a defined range of private-sector services~\cite{eaa2019}. 
The bodies that enforce accessibility and those that enforce consent operate in separate regulatory silos, an enforcement gap this paper documents empirically. 

\gls{wave}, produced by \gls{webaim}, and Google Lighthouse are the two automated accessibility testing tools. \gls{wave} analyses a rendered page for errors, contrast issues, alerts, and \gls{aria} landmarks, with a browser-extension variant able to evaluate dynamically generated content such as cookie notices~\cite{WAVEExtension}, while Lighthouse produces a weighted accessibility score out of 100 based on Axe checks incorporating \gls{wcag} rules~\cite{LighthouseAccessibilityScoring,Axecore2022}. The \gls{webaim} Million analysis of the top one million homepages, using \gls{wave}, found that 95.9\,\% contain detectable \gls{wcag} failures~\cite{webaim2025million}. A key limitation of both tools is that a high score does not guarantee a page is accessible in practice, since automated tools cannot test many \gls{wcag} criteria~\cite{frazao_comparing_2020}.

Evaluating website accessibility is possible through a range of automatic and manual methods, including screen readers and text-only browsers (e.g., \cite{southwell2013evaluation,schnellWebsitePrivacyNotification2021,ClarkeInvisible2024}).
Previous research highlights the need for accessible practices to protect this marginalised group, enabling appropriate security and privacy settings~\cite{inan2016internet}.
Self-perceptions can influence privacy and security, e.g., concealing a disability due to perceived stigma~\cite{hayesCooperativePrivacySecurity2019}. \gls{blv} users have also been reported to have difficulty completing security, with complications regarding, e.g., passwords and two-factor authentication~\cite{schmeelk2022digital}. 
While \gls{blv} users report awareness of online privacy and security risks~\cite{ahmedPrivacyConcernsBehaviors2015,napoli2021im}, their reliance on \glspl{at} disproportionately exposes them to deceptive design \cite{ClarkeInvisible2024}. Recent research reveals that accessibility barriers can unintentionally manifest as dark patterns, with \glspl{at} playing a complex dual role: although screen readers can expose certain hidden manipulations, they simultaneously exacerbate others through enforced sequential reading---a phenomenon termed \emph{assistive amplification}~\cite{lewis_inaccessible_2025,stanczyk_revealed_2026}. Given that compliance with accessibility frameworks like \gls{wcag} and the \gls{eaa} can serve as a direct regulatory lever~\cite{pellkvist_access_2026}, the distinct vulnerability of \gls{blv} users provides critical motivation for our large-scale investigation into the accessibility and risky properties of cookie notices.

Prior small-scale and practitioner evaluations show that cookie notices frequently suffer from contrast flaws, missing headings, delayed announcements, and poor keyboard support~\cite{ClarkeInvisible2024, marcus_herrmann_potentially_2023}. Furthermore, accessibility barriers on mobile interfaces can manifest as screen-reader-specific dark patterns such as focus hijacking and label deprivation~\cite{shi_when_2025}. Beyond consent interfaces, screen-reader navigation to build spatial awareness can inadvertently trigger third-party tracking on ad elements—a privacy burden absent for sighted users~\cite{amjad2026unintended}. Combined with inaccessible cookie notices, this reveals a pattern of compounding, involuntary tracking unique to assistive technology users. Yet, no prior work has conducted a large-scale automated analysis addressing both the accessibility and privacy implications of cookie notices for blind and low vision individuals.

\section{Methodology} \label{sec:cookie_methodology}
This section describes our website sample, crawler and screen reader instrumentation, cookie notice detection and interaction pipeline, and data processing methodology, before discussing the limitations of this approach.
\begin{figure}[t]
    \centering
    \includegraphics[width=0.8\linewidth,keepaspectratio]{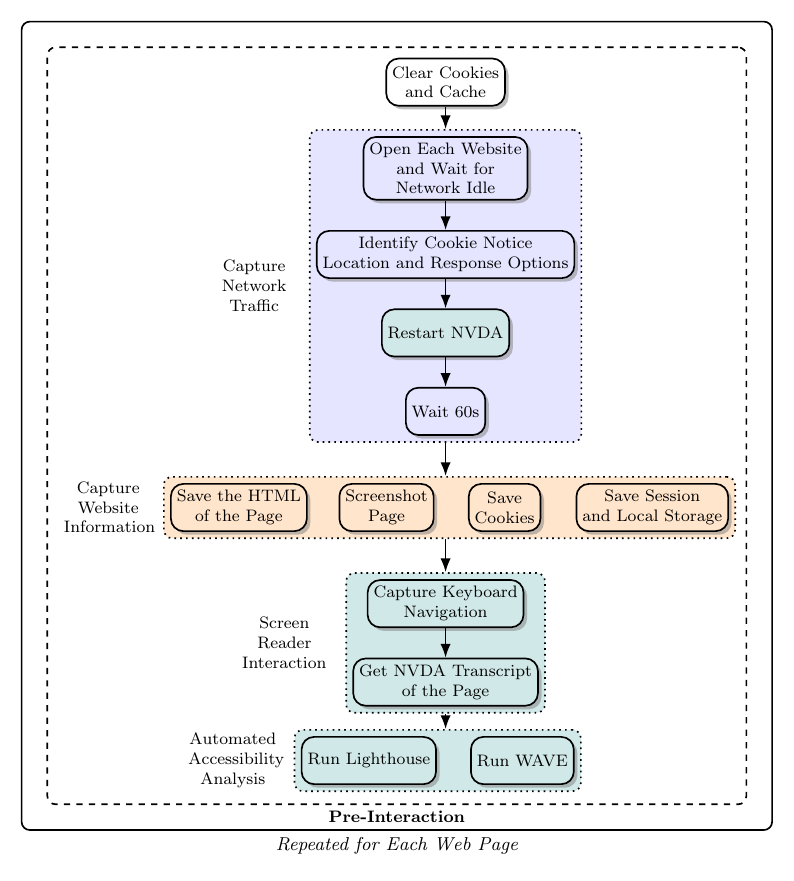}
    \caption{Pre-interaction flow of automated analysis.}
    \label{fig:pre-interaction}
\end{figure}
\begin{figure}[t]
    \centering
    \includegraphics[width=0.8\linewidth,keepaspectratio]{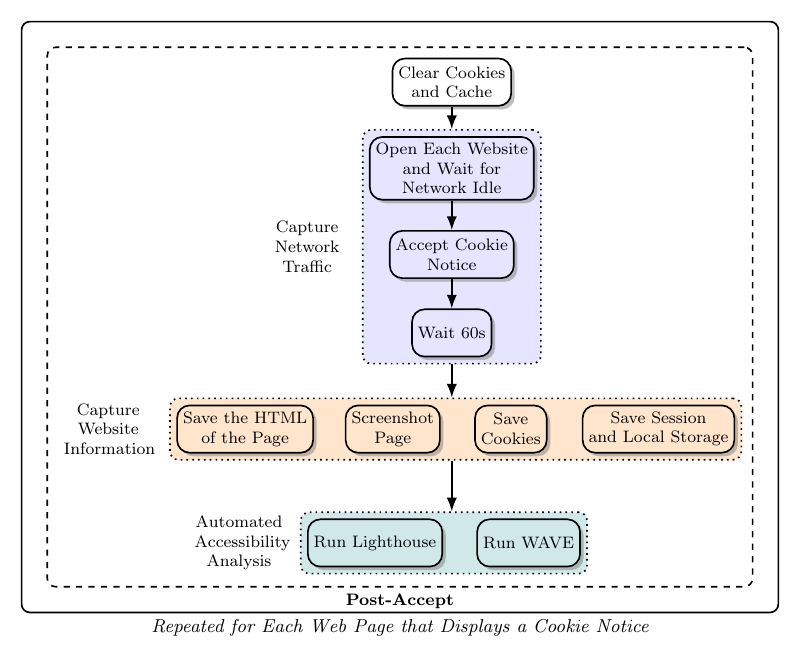}
    \caption{After-interaction flow when a cookie notice is accepted. An identical flow runs for cookie rejection as well.}
    \label{fig:accept}
\end{figure}
\subsection{Sample size} \label{sec:sample-size}
For our website selection, we use the Tranco list\footnote{Available at \url{https://tranco-list.eu/list/L76L4}.} ~\cite{LePochat2019} generated in March 2026.
This list has previously been used when examining the prevalence of both cookies and cookie notices~\cite{rasaii2023exploring,linBrowsingWithoutThirdParty2024, martinezLargeScaleWeb2025}. We focus on more popular websites as screen reader users often predominantly browse mainstream websites~\cite{rnibViLives2022}. Inspired by previous work~\cite{rasaii2023exploring,cassel2022omnicrawl}, we perform crawls across three distinct levels of website popularity: top 1k most popular websites (\emph{Top}), 1k websites randomly selected from the 1k-10K most popular tier (\emph{Middle}), and 1k websites randomly selected from the remaining sites (\emph{Bottom}).

Our sample size is a direct consequence of our chosen methodology: each site requires a full headed browser session with an \gls{nvda} restart per page (to clear the stale accessibility buffer, \S\ref{sec:screen-reader-interaction}) across three interaction flows, plus a 60-second dwell period per flow to capture deferred tracker requests, which together took a median of 163 seconds per site. The full 3,000-site, 3-tier crawl took approximately 12.4 days of wall-clock time in total, run sequentially from 5 to 18 May 2026. Those studies interact with the \gls{dom} directly and can therefore run many headless browser instances in parallel, whereas our methodology additionally requires a real, on-screen \gls{nvda} session per site; this per-site cost does not parallelise as easily as a headless, interaction-free crawl, so our sample cannot reach the 10,000--97,000-site scale of purely network-based audits~\cite{trevisanYearsEUCookie2019,nouwensDarkPatternsGDPR2020,bollinger2022automating,rasaii2023exploring,bouhoulaAutomatedLargeScale2023}.
Relative to the screen-reader-based cookie notice literature specifically, however, our sample is much larger than the largest prior (46 sites~\cite{ClarkeInvisible2024}; \autoref{tab:cookie_related_work}).

Following this methodology, our crawler also failed to load 939/3{,}000 (31\,\%) of sampled sites (\autoref{tab:errors}, \autoref{sec:unavailable-websites}). We navigated only over HTTPS, with no fallback to HTTP if the secure connection failed, and requested exactly the domain string given by the Tranco ranking, with no retry under a \texttt{www.} subdomain if the apex did not resolve. To establish how much these two gaps might explain the failure rate, we re-checked every site in the categories where a fallback could plausibly matter, DNS not resolved, TLS error, Connection failed, and Timed out, against live DNS and HTTP(S) requests in September 2026, roughly four months after the original crawl; hosting can change in that window, so this approximates rather than reconstructs conditions at crawl time.

Of the 498 DNS-not-resolved sites, 99 resolve under \texttt{www.} despite the bare apex still not resolving, and 38 of these serve live \texttt{HTTP 200} content, including sites we would otherwise have missed purely for lacking a \texttt{www.} retry (e.g.\ \texttt{meli.com}, \texttt{mufg.jp}, \texttt{pokemoncenter-online.com}). The remaining 384 resolve under neither hostname, consistent with the infrastructure, CDN, and ad-tech domains (e.g.\ \texttt{akamai.net}, \texttt{akadns.net}) that dominate this category. Of the 274 sites classified as a TLS error, a connection failure, or a timeout, 119 (43\,\%) respond over plain HTTP today, including established sites such as \texttt{delta.com}, \texttt{ge.com}, and \texttt{virginia.edu}. Altogether, 157/939 (16.7\,\%) of our recorded failures have live evidence that our missing HTTP and \texttt{www.} fallbacks caused an avoidable miss. HTTP 403 (bot detection, 99 sites) and HTTP 404 (39 sites) remain unrelated to protocol or subdomain choice. Redirects, meanwhile, were followed natively by our Chromium-based navigation rather than resolved only at the initial hostname, so redirect loops are captured under Connection failed rather than going undetected.

\subsection{Crawler} \label{sec:crawler}
We developed an automated measurement pipeline using Python and Playwright, operating a headed instance of Chrome to allow screen reader interaction. 
We cannot use a headless browser because \gls{nvda} hooks into the operating system's native accessibility APIs and provides an audio output of the on-screen application window. 
For each domain, we conducted three distinct interaction flows: (1) pre-interaction (no action taken), (2) post-accept, and (3) post-reject (\autoref{fig:pre-interaction} and \autoref{fig:accept}). To ensure independence, the browser state and cache were fully cleared between flows. For our privacy measurements, during each flow, we captured network traffic (including a 60-second dwell period for deferred trackers), HTML/CSS, cookies, and local/session storage. We integrated Google Lighthouse and WAVE to assess whether the notices and pages meet automated accessibility baselines. To capture the authentic BLV experience, we built a custom NVDA screen reader add-on. This add-on hooks into NVDA’s speech functions and virtual buffer, automatically exporting a timestamped transcript of the page's reading order and structural elements exactly as they are exposed to an AT user. %
Full implementation details are available in our open-source repository.

\subsection{Screen reader interaction} \label{sec:screen-reader-interaction}
We create a custom \gls{nvda} add-on to export a transcript of what \gls{nvda} would read aloud when interacting with the notice (similar to~\cite{southwell2013evaluation,ClarkeInvisible2024}). 
\gls{nvda} is a free, globally most popular, open-source Windows screen reader (estimated 1 million users~\cite{ScreenReaderUserSurvey}). 
\gls{nvda} add-ons are small pieces of downloadable code that extend or modify \gls{nvda}'s built-in functionality~\cite{momotaz2021understanding}. Our developed add-on is not distributed to end users and is used solely to capture speech output and traverse the accessible tree during our automated crawls.

We implement a file-based communication interface between a Playwright-driven crawler and \gls{nvda}, where the crawler issues commands (e.g., start, stop, navigate) and the add-on responds by capturing and exporting speech output. The add-on hooks into \gls{nvda}’s internal speech functions to record all spoken announcements with timestamps, enabling us to reconstruct the exact sequence of auditory feedback presented to the user. 
For more detailed analysis, the add-on can also actively traverse the page’s virtual buffer using \gls{nvda}’s navigation commands, extracting the reading order and key structural elements such as headings, links, landmarks, form fields, and images. 
\gls{nvda} uses a virtual buffer, an in-memory representation of the web page's accessible tree, for navigation. To clear and rebuild the buffer for each page, we need to restart \gls{nvda}, which clears the stale buffer from the previous page and forces a clean rebuild for the current one.
This approach allows us to approximate a real user’s interaction with the page, going beyond static \gls{dom} inspection to capture how content is actually exposed through \gls{at}. 
Focusing on the accessibility of the cookie notice, we perform the \gls{nvda} analysis only on the \emph{pre-interaction} version of each web page. 

\subsection{Cookie notice interaction and classification}
Our analysis automates cookie notice detection, location mapping, control options classification, and response execution. We base our implementation on BannerClick~\cite{rasaii2023cookiewall,rasaii2023exploring}, reusing its multilingual keyword vocabulary for cookie-notice detection and accept/reject-button identification. We extend BannerClick's detection algorithm with additional fallback passes to handle \glspl{cmp} that BannerClick's original Firefox/Selenium pipeline does not cover under our Chrome/Playwright setup, including notices rendered inside cross-origin iframes (e.g.\ TrustArc) and inside shadow \gls{dom} subtrees (e.g.\ OneTrust). If the automated approach cannot fully parse or classify the cookie notice or its response options, it flags the notice for manual review and classification.

\subsubsection{Identifying cookie notices and response options}
\label{sec:detection}
\paragraph{Detection and Interaction.}
We based our cookie notice detection and interaction on the BannerClick framework~\cite{rasaii2023exploring}, extending its multilingual keyword vocabulary. We enhanced BannerClick's detection algorithm with recursive iframe parsing and shadow DOM traversal to successfully identify complex CMPs (e.g., OneTrust, TrustArc) within our Playwright environment. Once detected, we utilised heuristic DOM analysis---evaluating visibility, z-index, and ARIA roles---to locate the notice bounding box. For interaction, we searched the notice subtree for accept, reject, or settings controls using BannerClick's vocabularies, supplemented with an exclusion filter to prevent false-positive clicks on partial-consent buttons (e.g., ``only necessary"). If a direct rejection button was absent but a settings route was available, the crawler attempted a two-step rejection flow. If the automated pipeline could not confidently detect or classify a notice, it was flagged for manual review.

\begin{figure}[t]
    \centering  
    \includegraphics[width=0.45\linewidth]{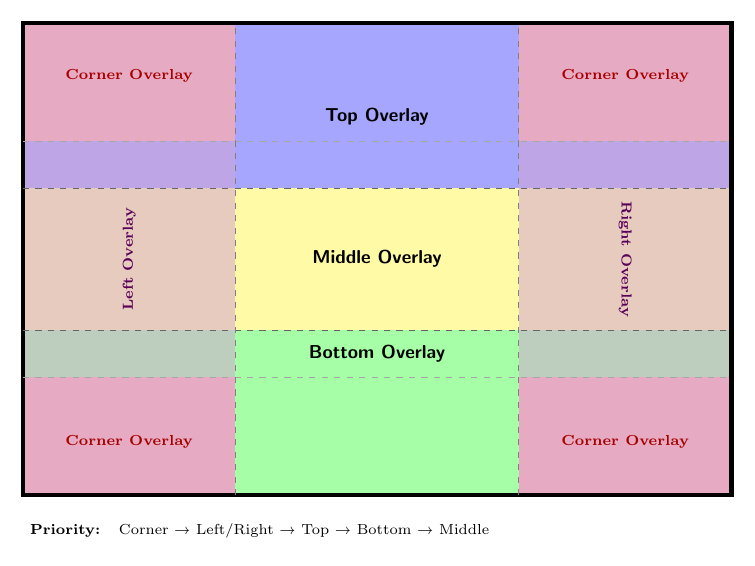}
    \caption[Positions for cookie notice location classification.]{Positions for cookie notice location classification; the location classification is taken from BannerClick~\cite{rasaii2023cookiewall,rasaii2023exploring}.}
    \label{fig:locations}
\end{figure}

\paragraph{Location.}
Once a cookie notice is detected, we record its bounding box relative to the viewport and assign it to one of the location categories (\autoref{fig:locations}), following the taxonomy of~\cite{rasaii2023cookiewall,rasaii2023exploring}:
(i) \emph{Top} or \emph{bottom} overlay: Spanning the full viewport width at either the top or the bottom.
(ii) \emph{Middle} overlay: Overlaid on the middle of the page content.
(iii) \emph{Left} or \emph{right} overlay: Anchored to the left or right edge of the viewport.
(iv) \emph{Corner} overlay: Anchored to the corners of the viewport.
(v) \emph{Full-page} overlay: That covers the underlying content entirely.

\paragraph{Response options.}
Following the response-option taxonomy of~\cite{rasaii2023cookiewall,rasaii2023exploring,mehrnezhadCrossPlatformEvaluationPrivacy2020}, we classify the set of interactive controls into one of six categories: 
(i) \emph{Accept only:} The only action is a positive consent button with no alternative.
(ii) \emph{Accept or settings:} Acceptance is offered alongside either a route to granular controls or granular controls embedded directly in the notice, but no direct reject option.
(iii) \emph{Accept or reject:} A direct binary choice of acceptance or rejection that does not require navigating to a separate settings panel.
(iv) \emph{Accept, reject, or settings:} Both a direct reject option and a settings route or panel are offered alongside acceptance.
(v) \emph{Accept or pay:} The user is offered a binary choice between accepting cookies or paying for an ad-free experience, with no option to reject tracking while retaining free access.
(vi) \emph{Informational:} The notice can only be dismissed without a meaningful consent action (e.g.\ ``OK'', ``Close'', ``Dismiss''), or where the notice is purely informational with no interactive elements.

\paragraph{Emphasis.}
To perform our compliance analysis, for notices offering more than one option, we identify which option is visually emphasised. Emphasis is determined by comparing the computed style of each interactive element with its siblings: we consider background colour saturation (filled vs \ outlined or ghost buttons), relative font weight and padding, and the element's position in the \gls{dom} order. A button is classified as emphasised if it is the only filled or coloured control, while the remaining options are rendered as secondary (outlined, text-only, or smaller) elements. If this cannot be classified automatically, we flag the notice for manual review. To our knowledge, this \gls{css}-based emphasis detection is a novel contribution: previous work (e.g.,~\cite{nouwensDarkPatternsGDPR2020}) manually inspect \gls{cmp} button designs and note that accept buttons are typically rendered more prominently than reject options.%

\subsubsection{Accepting and rejecting cookie notices}
\label{sec:interaction}
For the interaction, first, we relocate the banner using the same detection algorithm (\ref{sec:detection}), ensuring that the button search is confined to the notice element (not the full page). Next, we search the banner's subtree for a matching interactive element, collecting all elements matching the selector \texttt{button, a, [role=``button''], input[type=``submit''], input[type=``button'']}, and those nested inside shadow roots, and attempt \texttt{<button>} elements before other element types \cite{rasaii2023cookiewall,rasaii2023exploring}. 
Following each click, we verify that the interaction succeeded by checking that the banner element is no longer visible in the viewport. If this check fails, we flag the notice for manual review, as the interaction may have failed or triggered a non-standard response (e.g., a cookie settings panel or a two-step rejection flow).

For \emph{acceptance}, we match buttons whose visible text or \texttt{aria-label} contains one word from BannerClick's accept-word list (e.g.\ \emph{accept}, \emph{agree}, \emph{allow})~\cite{rasaii2023cookiewall,rasaii2023exploring}. We then apply an unacceptable filter, discarding any button whose text also contains a qualifying exclusion term (e.g.\ \emph{only}, \emph{without}, \emph{just}), which prevents false positives such as ``only necessary cookies'' from being treated as a full-consent action. Unlike BannerClick~\cite{rasaii2023exploring}, which documents clicking ``NOT Accept'' as a known failure mode affecting 13 of 444 sites precisely because no such filter is applied, we additionally apply this exclusion filter as a novel extension to prevent that class of false positives. We click the first button that passes both criteria (\autoref{fig:accept}). 

For \emph{rejection}, we apply the same banner localisation and button collection procedure but match against BannerClick's reject-word list (e.g.\ \emph{reject}, \emph{decline}, \emph{refuse}, \emph{deny}). 
Similarly, if the banner remains after a direct reject attempt and the notice offers a settings route, we attempt a two-step rejection: we first click a settings or preferences button drawn from BannerClick's settings-word list, wait for the resulting panel to render, and then search again for a confirm-selection or reject-all control. For notices that have a setting or reject button but neither a direct nor a two-step rejection succeeds, we flag them for manual verification (\autoref{fig:accept}).

\subsection{Tracking and cookie classification} \label{sec:tracking-cookie-classification}
Due to the volume of results, we create an automated \emph{Go} script to post-process all collected data.
To classify network traffic, we cross-referenced all HTTP requests against EasyPrivacy~\cite{EasyPrivacy}, EasyList without adult filter~\cite{EasyList}, AdGuard's Tracking Protection Filter~\cite{AdGuardTracking}, AdGuard's Base Filter~\cite{AdGuardBase}, and AdGuard's Mobile Ads Filter~\cite{AdGuardMobile}, defining matching requests as trackers. Third-party requests were identified by comparing request domains against the first-party site domain. Cookies and client-side storage were categorised (e.g., Functional, Marketing, Analytics) using the Open Cookie Database and the Cookie Database API. 

\begin{table*}[t]\centering\footnotesize
\caption{Cookie notice control options and UK GDPR/PECR compliance.}\label{tab:scale_options}
\begin{threeparttable}
\begin{tabularx}{\linewidth}{Xcrrr} \toprule
  \textbf{Control options} & \makecell{\textbf{Emphasised}\\ \textbf{option}} & \textbf{Sites} & \textbf{\%} & \makecell{\textbf{\gls{ukgdpr}/\gls{pecr}} \\ \textbf{violation}$^{\dag}$} \\ \midrule
  Accept, Reject, or Settings & Accept & 121 & 6\,\% & \textbf{Yes}$^{e}$ \\
  Accept, Reject, or Settings & Equal & 422 & 20\,\% & No \\
  Accept or Settings & Accept & 135 & 7\,\% & \textbf{Yes}$^{a}$ \\
  Accept or Settings & Equal & 101 & 5\,\% & \textbf{Yes}$^{a}$ \\
  Accept or Reject & Accept & 63 & 3\,\% & \textbf{Yes}$^{e}$ \\
  Accept or Reject & Equal & 106 & 5\,\% & No \\
  Accept only & Accept & 29 & 1\,\% & \textbf{Yes}$^{b}$ \\
  Accept or Pay & Accept & 5 & 0\,\% & \textbf{Yes}$^{c}$ \\
  Accept or Pay & Equal & 20 & 1\,\% & \textbf{Yes}$^{c}$ \\
  Informational only & Other & 56 & 3\,\% & \textbf{Yes}$^{b}$ \\
  No Notice&NA & 1{,}003 & 49\,\% & \textbf{Yes}$^{d}$ \\
  \bottomrule
\end{tabularx}
\begin{tablenotes}\footnotesize
  \item[$\dag$] Non-compliant if no equally accessible reject path, per \gls{ukgdpr} Art.~4(11) \&
  7(3)~\cite{UKGDPR}.
  \item[$a$] Reject requires extra steps; violates withdrawal-symmetry
  requirement~\cite{informationcommissionersofficeGuidanceUseCookies2022,UKGDPR}.
  \item[$b$] No refusal mechanism; no lawful basis under \gls{ukgdpr} Art.~6~\cite{UKGDPR}.
  \item[$c$] Payment-gated consent not freely given under \gls{ukgdpr} Art.~7(4)~\cite{UKGDPR} and \gls{duaa}~\cite{informationcommissionersofficeGuidanceUseCookies2022, ICOAcceptOrPay}.
  \item[$d$] No consent mechanism; violates \gls{pecr} Reg.~6~\cite{PECR} \&
  \gls{ukgdpr} Art.~6~\cite{UKGDPR}.
  \item[$e$] Reject path present but accept button visually emphasised; asymmetric presentation may invalidate freely-given consent under \gls{ukgdpr} Art.~4(11) \& 7(3)~\cite{UKGDPR,ICOAcceptOrPay}.
\end{tablenotes}
\end{threeparttable}
\end{table*}
\subsection{Accessibility evaluation} \label{sec:accessibility-evaluation}
To identify issues directly attributable to cookie notices, we used a differential analysis method, comparing Lighthouse and WAVE scores across the three interaction phases to isolate regressions caused by notice dismissal. Adapted from~\cite{ClarkeInvisible2024}, we also evaluated the NVDA transcripts against seven screen reader accessibility metrics, assigning a pass/fail to evaluate how the notice was announced, titled, and navigated. These metrics are: Readable, Immediately Read, Keyboard Navigable, Link or Button Purpose, Abbreviations Explained, Page Titled, and Cookie Notice Titled; defined extensively in~\autoref{appendix:cookie-accessibility-metrics}.

\subsection{Limitations}\label{sec:limitations}
Because our methodology is entirely automated, our metrics serve as scalable proxies evaluating whether the banner meets automated accessibility baselines, but cannot capture the nuanced navigation strategies, cognitive load, or subjective experiences of real \gls{blv} users. Furthermore, our accessibility evaluation is limited to desktop browsing using \gls{nvda}, meaning interaction behaviours may differ on mobile devices or alternative \glspl{at} such as \gls{jaws} and VoiceOver.

Finally, consistent with prior large-scale measurements,
we evaluated only the landing pages of a global Tranco list (generated in March 2026) during a single crawl snapshot run 5--18 May 2026. Consequently, our findings may not capture intra-site variations, the continuously evolving nature of web tracking, or guarantee that all lower-tier sites in our sample are strictly subject to \gls{ukgdpr} and \gls{pecr}~\cite{rnibViLives2022}. Future user studies and multi-device evaluations are required to validate and complement our work.

\begin{table*}[t]
  \centering
  \scriptsize
  \setlength{\tabcolsep}{3pt}
  \caption{Tracker request counts per site by phase.}
  \label{tab:tracker-request-summary}
  \begin{tabular}{lrrrr}
    \toprule \textbf{Metric} & \textbf{No notice} & \textbf{Pre} & \textbf{Post-acc} & \textbf{Post-rej} \\ \midrule
    Avg.\ tracker reqs/site & 35.9 & 43.4 & 218.5 & 44.5 \\
    Median tracker reqs/site & 7 & 16 & 51 & 18 \\
    Mode tracker reqs/site & 0 & 0 & 16 & 16 \\
    Total tracker reqs & 35{,}971 & 45{,}892 & 196{,}466 & 26{,}795 \\
    Total network reqs & 155{,}806 & 209{,}157 & 363{,}301 & 121{,}985 \\
    Tracker rate & 23.1\,\% & 21.9\,\% & 54.1\,\% & 22.0\,\% \\
    \bottomrule
  \end{tabular}
\end{table*}
\begin{table*}[t]\setlength{\tabcolsep}{3pt}
  \centering
  \footnotesize
  \caption{Differences in tracker request counts across phases.}
  \label{tab:tracker-differences}
  \begin{tabular}{lrrr}
    \toprule
    \textbf{Comparison} & \textbf{More} & \textbf{Same} & \textbf{Fewer} \\ \midrule
    Post-accept vs\ pre-interaction (899 sites) & 774 & 66 & 59 \\
    Post-reject vs\ pre-interaction (602 sites) & 375 & 118 & 109 \\
    Post-reject vs\ post-accept (590 sites) & 38 & 101 & 451 \\
    \bottomrule
  \end{tabular}
\end{table*}

\section{Privacy results} \label{sec:privacy-results}
This section addresses RQ1 by reporting on tracking behaviour across consent phases, response options and compliance.
Out of the 3,000 sampled sites, 939 failed to load
(with over half being DNS or TLS errors rather than possible bot detection, as detailed in Appendix~\ref{sec:unavailable-websites}), 1,003 had no notice, and 1,058 had a cookie notice. The low rate of HTTP 403 blocks across all Tranco tiers confirms our crawler was not systematically locked out of secure sites. Therefore, our subsequent analysis focuses on the 2,061 successfully reachable domains.

\subsection{Cookie notice classification}
A cookie notice was detected on 1,058 of the 2,061 reachable sites (51\,\%), successfully accepted on 899 of those (85\,\%) (automatic: 861, manual: 38).
Of the 718 sites where a reject or decline button was identified, it was successfully clicked on 602 (84\,\%) (automatic: 594, manual: 8). Automatically, we classified the position and response options of 745 notices and identified 954 sites without a cookie notice, while manually classifying 362 sites. 

\subsubsection{Position}
Cookie notices most commonly appeared as a \emph{bottom overlay} (460 of 1,058 sites, 43.5\,\%), followed by a \emph{middle overlay} (33.7\,\%), a \emph{corner overlay} (123 sites, 11.6\,\%), and \emph{top overlay} (9\,\%). The least common positions were \emph{left, right, and full-page overlay} (<2\,\%), where the notice covers the entire page and requires scrolling to access the underlying content. 

Previous research found that placing a consent notice in the lower portion of the screen increases interaction rates~\cite{utzInformedConsentStudying2019}, and that barrier-style (middle overlay) notices are ignored 3.6 times less often than banner-style notices. 
For \gls{blv} users, position has a direct effect on screen reader reading order: a bottom overlay may be placed late in the \gls{dom} and may be encountered only after a screen reader user has traversed the entire page content above it, whereas a top overlay or a notice injected as the first focusable element is announced early. We discuss this in (\autoref{sec:accessibility-results}). 

\subsubsection{Control types}
The most common control type was \emph{accept, reject, or settings} (543 sites, 51\,\%), where there was an accept button, a reject button, and a route to granular cookie settings. 
The next most common was \emph{accept or settings} (22\,\%), where the notice offered an accept button and a route to granular settings, but no direct reject button. Next was \emph{accept or reject} (16\,\%), where the notice offered direct accept and reject buttons but no route to granular settings. Other control types included \emph{informational only} (5\,\%) offering no meaningful consent options, \emph{accept or pay} (2\,\%) offering a choice between accepting cookies or paying for an ad-free experience, and \emph{accept only} (3\,\%) with only a direct accept button and no alternative.

Button labels were extracted from the pre-interaction HTML of the 1,058 cookie-notice pages using the same keyword matching logic as the scanner. Accept-style labels were identified on 817 pages and reject-style labels on 600 pages. 
The most common labels (\autoref{tab:labels}) were ``Accept all'' (211 sites, 20\,\%), and ``Reject all'' (222 sites, 21\,\%). However, a long tail of less common labels was observed for both accept and reject buttons, with 147 unique accept labels and 112 unique reject labels across the dataset. 
\begin{figure}[t]
  \centering
  \includegraphics[width=0.8\linewidth,keepaspectratio]{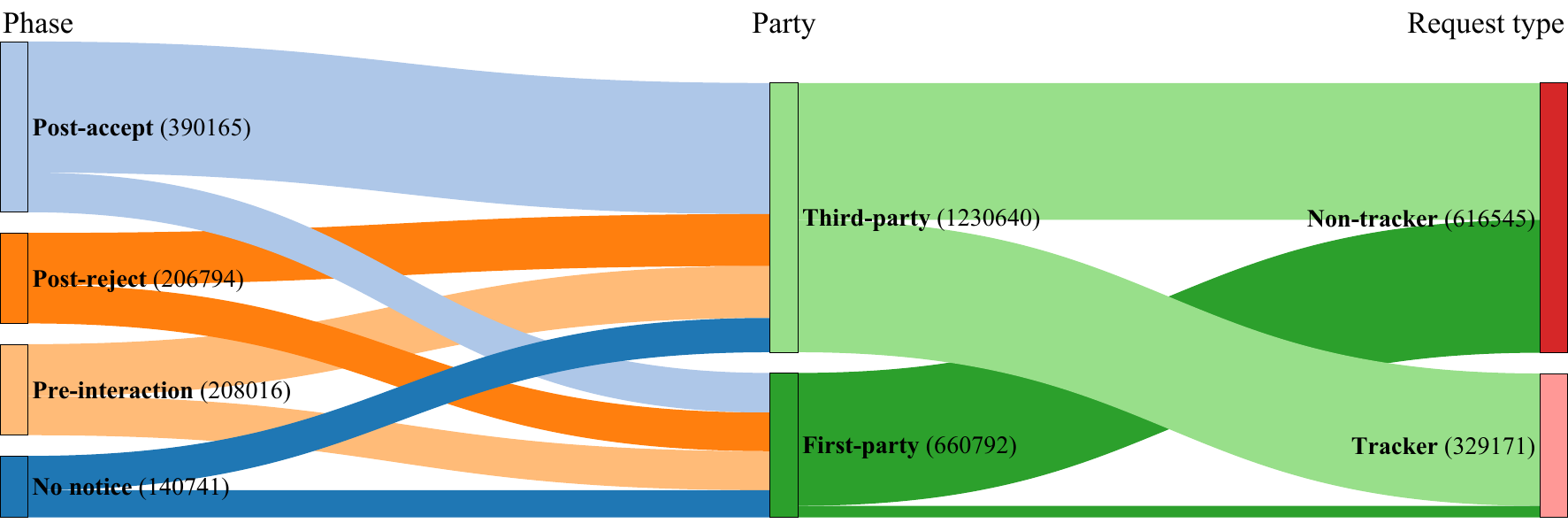}
  \caption{The flow of requests across phases.}
  \label{fig:sankey_trackers}
\end{figure}
\begin{figure}[t]
  \centering
  \includegraphics[width=0.8\linewidth,keepaspectratio]{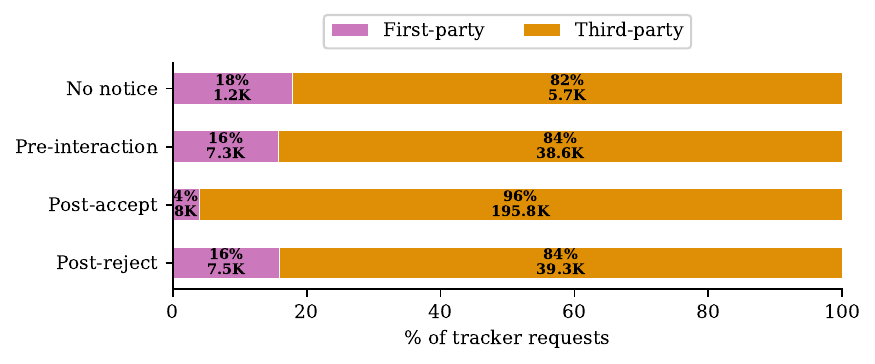}
  \caption{Distribution of tracker requests across phases.}
  \label{fig:tracker_party}
\end{figure}

\subsubsection{Emphasised option}
Among notices that offer a choice (1,002), 649 (65\,\%) presented the accept and reject or settings options with \emph{equal} visual weight, a positive finding, with no deliberate dark pattern.
However, 353 notices (35\,\%) emphasised the accept option, typically by rendering it as a filled or coloured button, while the reject or settings options were outlined, text-only, or smaller, a nudge pattern previously identified in the top five consent management platforms~\cite{nouwensDarkPatternsGDPR2020}.   
The remaining 56 notices (5\,\%) were classified as \emph{not applicable} because they offered no choice (informational only) and therefore had only a close, okay, or dismiss button. 
Under \gls{ukgdpr} Article~4(11), consent must be ``freely given'', meaning the user must have a genuine choice with no detriment for refusing~\cite{UKGDPR,informationcommissionersofficeGuidanceUseCookies2022}. 

\subsubsection{UK GDPR/PECR compliance}
\autoref{tab:scale_options} classifies all reachable sites by the control options offered to visitors, using the taxonomy described in~\S\ref{sec:detection}. The {\gls{ukgdpr}/\gls{pecr} violation} column indicates whether the notice design satisfies the \gls{ukgdpr} requirement for freely given, unambiguous consent.
Of the 1,058 sites that displayed a cookie notice, 528 (50\,\%) are compliant, and 530 (50\,\%) are not. 
1,003 sites (49 \%) displayed no cookie notice at all. While some of these may be outside the territorial scope of UK GDPR/PECR, or rely on essential cookies, any of these domains actively targeting or monitoring UK users are doing so in violation of the law.

This violation rate is substantially lower than the 90.2\,\% reported for 97k EU websites by~\cite{bouhoulaAutomatedLargeScale2023} and the 94.7\,\% reported across almost 30k European websites with \glspl{cmp} by~\cite{bollinger2022automating}. This gap can be attributed to methodology rather than to an underlying improvement in compliance: we flag violations detectable from notice structure and visual design (absent reject paths, accept-only, accept-or-pay, and emphasised accept buttons), whereas this prior work detects mismatches between declared and observed cookie behaviour through NLP-based banner interaction~\cite{bouhoulaAutomatedLargeScale2023} or ML-based cookie-purpose classification~\cite{bollinger2022automating}. Our 50\,\% should therefore be read as a lower bound on the true violation rate.

\begin{figure}[t]
  \centering
  \includegraphics[width=0.8\linewidth,keepaspectratio]{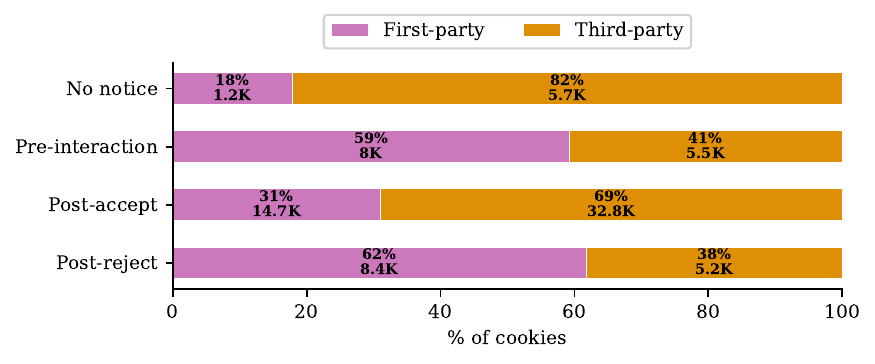}
  \caption{Distribution of cookies across phases.}
  \label{fig:cookie_party}
\end{figure}
\begin{figure}[t]
  \centering
  \includegraphics[width=0.8\linewidth,keepaspectratio]{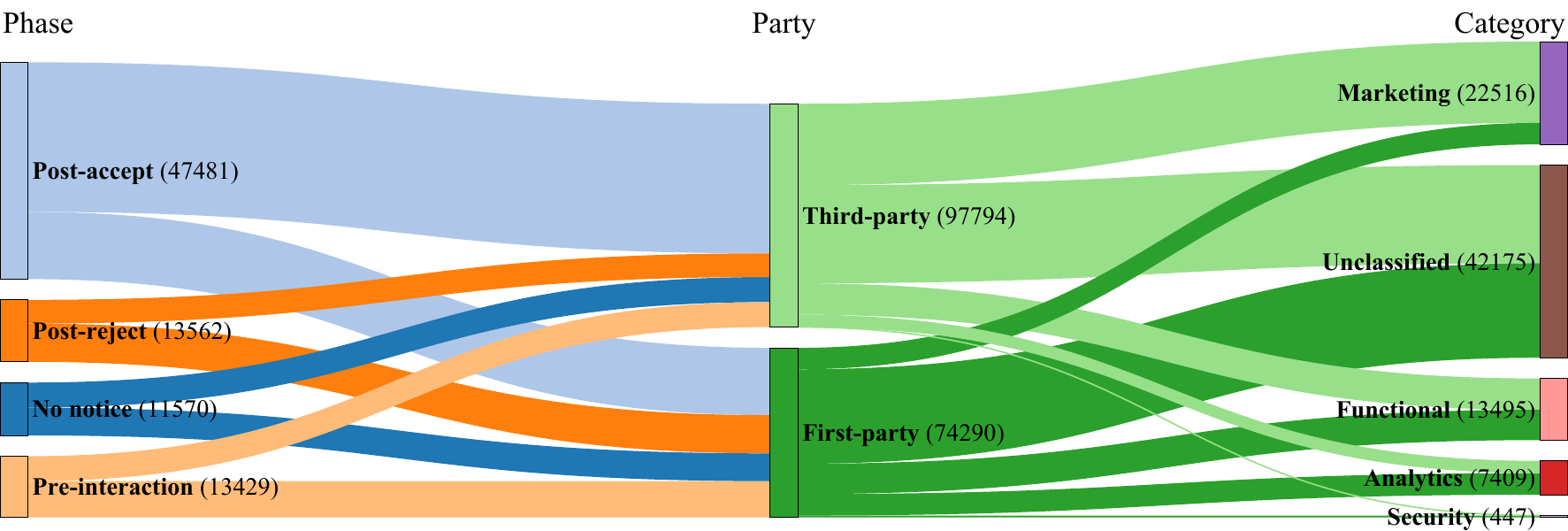}
  \caption{The flow of cookies across phases.}
  \label{fig:sankey_cookies}
\end{figure}
\subsection{Network traffic analysis}
Across all sites, we recorded 983,509 requests, of which 850,249 fall into the four clean categories shown in \autoref{tab:tracker-request-summary} (no-notice, pre-interaction, successfully-accepted, and successfully-rejected scans); the remainder comes from the 939 sites that failed to load, sites where notice detection was a false positive, and post-interaction phases on scans without a confirmed successful accept or reject. As shown in Tables \ref{tab:tracker-request-summary} and \ref{tab:tracker-differences}, sites with a cookie notice averaged 43.4 tracker requests pre-interaction, which surged to 218.5 post-accept---a five-fold increase demonstrating the substantial tracking cost of consent. Conversely, rejection limited post-interaction trackers to 44.5 on average, barely above the pre-interaction baseline. However, pre-interaction and post-reject tracking remains non-trivial: of the 602 rejected sites, 62\,\% still saw increased tracking compared to baseline. 
Additionally, \autoref{fig:sankey_trackers} (and \autoref{fig:tracker_party}) show how the distribution of first- and third-party trackers changes: third-party trackers spiked from 84\,\% (pre-interaction) to 96\,\% (post-accept), heavily driven by ad-networks (across all phases) like \texttt{googlesyndication.com}, \texttt{doubleclick.net}, \texttt{google.com}, \texttt{taboola.com}, and \texttt{googletag-manager.com}, together accounting for 24\,\% of all third-party tracker requests.

These results align with prior work: \cite{bouhoulaAutomatedLargeScale2023} find that 65.4\,\% of sites offering a reject option still deploy analytics or advertising cookies after explicit negative consent, and \cite{bollinger2022automating} find that 21.3\,\% of sites create cookies despite a negative consent signal. Our finding that 62\,\% of sites with a completed rejection show more tracker requests than the pre-interaction baseline is consistent with this pattern of persistent post-rejection tracking. Another work \cite{ormeno_cookie_2024} similarly reports that rejection substantially limits, but does not eliminate, tracker activity relative to acceptance.

\subsection{Browser storage analysis}
We analyse the client-side storage observed during each scan phase (\autoref{tab:storage_master}) with a focus on cookies. 
After rejecting the cookie notice, the average was 12.8 cookies per site, with a total of 7,695 cookies across 602 sites, suggesting rejection does not fully prevent cookie setting.
Sites without a notice averaged 12.4 cookies, matching the pre-interaction baseline of sites with a notice (12.7). Acceptance triggered a four-fold increase to 50.6 cookies per site, whereas rejection limited the average to 12.8, indicating that rejection mitigates but does not fully prevent cookie setting. Cookie composition shifts significantly by phase (\autoref{tab:cookie_cats}): Marketing cookies double post-accept (17 \% to 34 \%), while Functional categories drop (23 \% to 12 \%). Post-reject, Marketing falls to 13 \% and Functional rises to 30\%, with Unclassified cookies remaining stable near 50\%. Notably, sites without a notice exhibit higher unconsented Analytics tracking (16 \%) than pre-interaction sites (9 \%).

\begin{table*}[h]\setlength{\tabcolsep}{3pt}
  \centering
  \footnotesize
  \caption{Client-side storage counts by type and phase.}\label{tab:storage_master}
  \begin{threeparttable}
    \begin{tabularx}{\textwidth}{lXXXXXXXXXXXX}%
    \toprule
    \multirow{2}{*}{\textbf{Type}} & \multicolumn{3}{c}{\textbf{No notice}} & \multicolumn{3}{c}{\textbf{Pre-interaction}} & \multicolumn{3}{c}{\textbf{Post-accept}} & \multicolumn{3}{c}{\textbf{Post-reject}} \\ \cmidrule(lr){2-4}\cmidrule(lr){5-7}\cmidrule(lr){8-10}\cmidrule(lr){11-13}
    & \textbf{Total} & \textbf{Sites} & \textbf{Avg} & \textbf{Total} & \textbf{Sites} & \textbf{Avg} & \textbf{Total} & \textbf{Sites} & \textbf{Avg} & \textbf{Total} & \textbf{Sites} & \textbf{Avg} \\ \midrule
    Cookies          & 12{,}478 & 1{,}003 & 12.4 & 13{,}429 & 1{,}058 & 12.7 & 45{,}507 & 899 & 50.6 & 7{,}695 & 602 & 12.8 \\
    Local storage    & 6{,}130  & 609  & 10.1 & 6{,}466  & 877  & 7.4  & 12{,}083 & 717 & 16.9 & 3{,}652 & 447 & 8.2  \\
    Session storage  & 1{,}567  & 609  & 2.6  & 2{,}414  & 877  & 2.8  & 3{,}290  & 717 & 4.6  & 1{,}429 & 447 & 3.2  \\
    \bottomrule
  \end{tabularx}
  \begin{tablenotes}\footnotesize
  \item Avg = average per site in the phase. For cookies, Sites = all sites in that phase (including those setting zero cookies). For local/session storage, Sites = sites for which storage data was recorded.
  \end{tablenotes}
  \end{threeparttable}
\end{table*}

Further analysis (\autoref{fig:cookie_party}) shows that in pre-interaction and post-reject, the majority of cookies are first-party (59\,\% and 62\,\% respectively), while in post-accept, the majority are third-party (69\,\%). When there is no notice, 53\,\% of the cookies are first-party, and 47\,\% are third-party. Therefore, accepting the notice leads to a substantial increase in third-party cookies, whereas rejecting it may reduce them but does not eliminate them; some sites may still set cookies even after rejection. 
This distribution changes from pre-interaction to post-accept and post-reject (\autoref{fig:sankey_cookies}).
\begin{figure}[t]
  \centering
\includegraphics[width=0.8\linewidth,keepaspectratio]{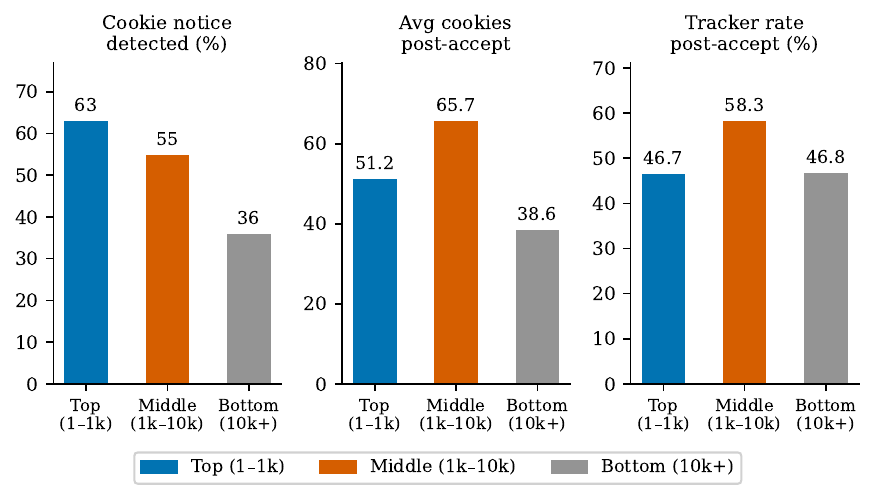}
  \caption[Cookie notice metrics by tier]{Cookie notice detection, automated rejection success, and post-accept/post-reject cookie and tracker rates across the three tiers.}
  \label{fig:tier_metrics}
\end{figure}
\subsection{Cross website tier comparison}
Comparing sites by Tranco tier (\autoref{fig:tier_metrics}) reveals that notice prevalence and automated rejection success decline sharply with popularity (Top: 63 \% prevalence and 79 \% rejection success; Bottom: 36 \% and 36 \%), reflecting lower standardisation on less prominent domains. Paradoxically, Middle-tier sites deploy the most aggressive tracking ecosystems, exhibiting the highest tracker rates both pre-interaction (26.7\,\%) and post-accept (60.6\,\%), alongside the highest post-accept cookie load (avg.\ 61.2), exceeding Top-tier domains. Conversely, while Bottom-tier sites carry the lowest baseline cookie loads, they average 34.3 cookies post-rejection---the highest of any tier---indicating that consent refusal mechanisms are least effective on lower-ranked websites. Full analysis is provided in \autoref{AppD}.
\begin{table*}[t]
  \centering
  \footnotesize
  \caption{Number of cookie categories by scan phase.}
  \label{tab:cookie_cats}
  \begin{tabular}{lrrrrrrrrrrrr}
    \toprule \textbf{Category} & \multicolumn{3}{c}{\textbf{No notice}} & \multicolumn{3}{c}{\textbf{Pre-interaction}} & \multicolumn{3}{c}{\textbf{Post-accept}} & \multicolumn{3}{c}{\textbf{Post-reject}} \\
    \cmidrule(lr){2-4}\cmidrule(lr){5-7}\cmidrule(lr){8-10}\cmidrule(lr){11-13}
    & \textbf{$n$} & \textbf{Sites} & \textbf{\%} & \textbf{$n$} & \textbf{Sites} & \textbf{\%} & \textbf{$n$} & \textbf{Sites} & \textbf{\%} & \textbf{$n$} & \textbf{Sites} & \textbf{\%} \\ \midrule
      Unclassified & 6{,}183 & 713 & 50 & 6{,}737 & 909 & 50 & 22{,}008 & 870 & 48 & 3{,}882 & 523 & 50 \\
      Marketing & 2{,}586 & 421 & 21 & 2{,}311 & 465 & 17 & 15{,}291 & 741 & 34 & 1{,}017 & 245 & 13 \\
      Functional & 1{,}620 & 525 & 13 & 3{,}039 & 860 & 23 & 5{,}279 & 818 & 12 & 2{,}289 & 529 & 30 \\
      Analytics & 1{,}988 & 536 & 16 & 1{,}273 & 430 & 9 & 2{,}691 & 720 & 6 & 440 & 175 & 6 \\
      Security & 84 & 66 & 1 & 55 & 47 & 0 & 212 & 200 & 0 & 65 & 60 & 1 \\
      \makecell[l]{Statistics\\(anonymous)} & 17 & 17 & 0 & 14 & 13 & 0 & 26 & 24 & 0 & 2 & 2 & 0 \\
    \midrule
    \textbf{Total} & \textbf{12{,}478} & 785 & & \textbf{13{,}429} & \textbf{1{,}021} & & \textbf{45{,}507} & 898 & & \textbf{7{,}695} & 595 & \\
    \bottomrule
  \end{tabular}
\end{table*}
\section{Accessibility results} \label{sec:accessibility-results}
This section addresses RQ2, examining how the presence and interaction with a cookie notice affects \gls{wcag} 2.2 accessibility scores and how cookie notices are accessible to screen reader users.

\subsection{Automated accessibility testing}
To evaluate notice impacts, Lighthouse and \gls{wave} tools ran across pre-interaction, post-accept, and post-reject phases. Among 899 accepted sites, 892 yielded valid pre- and post-accept Lighthouse scores, while 7 failed due to timeouts, exits, or missing categories. WAVE measurements failed on 7 sites, with only one site missing both tools' scores, resulting in 13 distinct sites with missing data. Out of 602 rejection sites, 597 achieved valid Lighthouse scores, and 598 had valid WAVE scores; as no site was missing both tools' scores here, 9 distinct sites failed in total across the two tools.
As reported in \autoref{tab:accessibility_with_cookie_notices}, sites \emph{without} a cookie notice (1,003 sites) had a mean pre-interaction Lighthouse score of 82.6 (vs. 88.5 for sites with a notice). \gls{wave} errors averaged 18.5 (no notice) vs 18.7 (with notice); contrast errors 31.1 vs 27.5; and alerts 60.2 vs 90.7. 
\begin{table*}[t]\centering\footnotesize
  \caption{WAVE and Lighthouse accessibility metrics by phase.}
  \label{tab:accessibility_with_cookie_notices}
  \begin{threeparttable}
  \setlength{\tabcolsep}{3pt}
  \begin{tabular}{lrrrr}
    \toprule \textbf{Metric} & \makecell{\textbf{No notice}} & \makecell{\textbf{Pre-interaction}} & \makecell{\textbf{Post-accept}} & \makecell{\textbf{Post-reject}} \\ \midrule
    Lighthouse score (avg)     & 82.6 & 88.5 & 88.3 & 89.6 \\
    WAVE errors (avg)          & 18.5 & 18.7 & 15.8 & 13.6 \\
    \quad Median & 4 & 6 & 6 & 4 \\
    \quad Mode   & 0 & 0 & 0 & 0 \\
    WAVE contrast errors (avg) & 31.1 & 27.5 & 24.9 & 19.9 \\
    \quad Median & 8 & 9 & 8 & 5 \\
    \quad Mode   & 0 & 0 & 0 & 0 \\
    WAVE alerts (avg)          & 60.2 & 90.7 & 72.1 & 67.6 \\
    \quad Median & 18 & 34 & 29 & 25 \\
    \quad Mode   & 2 & 1 & 1 & 1 \\
    \bottomrule
  \end{tabular}
  \begin{tablenotes}\footnotesize
  \item LH\,=\,Lighthouse score (0--100); higher is better. \gls{wave} metrics: lower is better.
  \end{tablenotes}
  \end{threeparttable}
\end{table*}
\begin{table*}[t]\centering\footnotesize
  \caption[Comparing Lighthouse and WAVE scores before vs.\ after accepting or rejecting the cookie notice.]{Comparing Lighthouse and WAVE scores before vs.\ after accepting or rejecting the cookie notice. $\Delta$ is the change relative to the pre-interaction baseline; $p$/$r$ are from paired Wilcoxon signed-rank tests, testing whether the notice's dismissal significantly changes page accessibility.}\label{tab:a11y}
  \begin{threeparttable}
  \begin{tabular}{lrrrllrrll} \toprule
    \textbf{Metric} & \textbf{Pre} & \textbf{Post-acc} & $\Delta$\,\textbf{Acc} & $p_\text{acc}$ & $r_\text{acc}$ & \textbf{Post-rej} & $\Delta$\,\textbf{Rej} & $p_\text{rej}$ & $r_\text{rej}$ \\ \midrule
    Lighthouse score    & 89.3 & 88.3 & \textbf{-1.0} & $<\!0.001$ & $-0.43$ & 89.6 & \textbf{+0.3} & $<\!0.001$ & $-0.40$ \\
    WAVE errors         & 15.1 & 15.8 & \textbf{+0.7} & $<\!0.001$ & $+0.51$ & 13.7 & \textbf{-1.4} & $0.145$ & $-0.18$ \\
    WAVE contrast errs  & 24.9 & 24.4 & \textbf{-0.5} & $<\!0.001$ & $-0.58$ & 19.1 & \textbf{-5.8} & $<\!0.001$ & $-0.62$ \\
    WAVE alerts         & 80.0 & 70.3 & \textbf{-9.7} & $0.059$ & $-0.11$ & 64.6 & \textbf{-15.4} & $<\!0.001$ & $-0.32$ \\
    \bottomrule\end{tabular}
  \begin{tablenotes}\footnotesize
  \item LH\,=\,Lighthouse score (0--100); higher is better. \gls{wave} metrics: lower is better. $p$-values: two-sided Wilcoxon signed-rank test (pre vs\ post); $r$: rank-biserial correlation (effect size). Post-reject figures are over the 597 sites with paired scores for both phases; the all-valid-measurement post-reject \gls{wave} error mean (598 sites) is 13.6, with the difference from the 13.7 shown here being a rounding artefact of the smaller paired sample.
  \end{tablenotes}
  \end{threeparttable}
\end{table*}

\begin{figure}[b]
  \centering
  \includegraphics[width=0.6\linewidth]{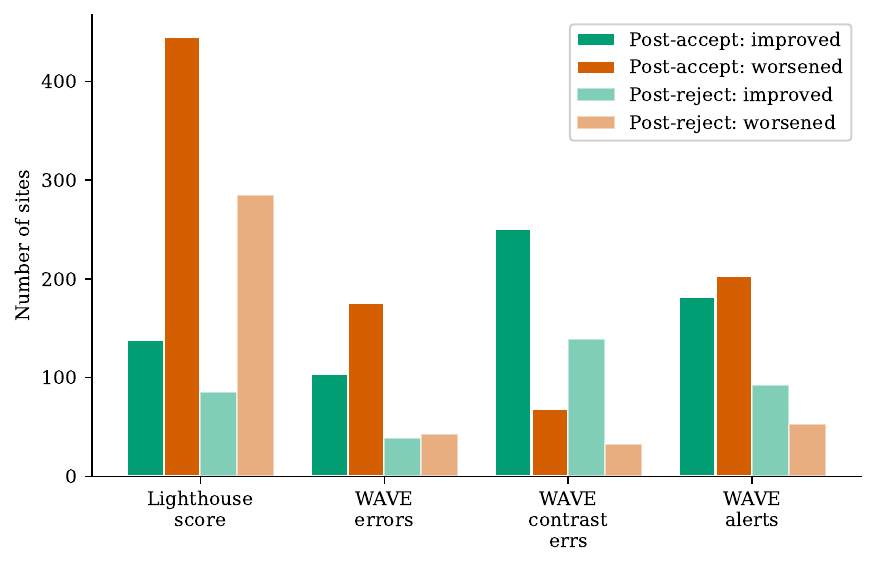}
  \caption{Number of sites improved or worsened per accessibility metric after accepting or rejecting the cookie notice.}
  \label{fig:a11y_counts}
\end{figure}

\autoref{fig:a11y_counts} summarises the number of sites that improved or worsened per metric, and \autoref{tab:a11y} compares accessibility metrics before and after accepting the cookie notice for the 892 sites where both pre-interaction and post-accept Lighthouse scores were available. Post-reject metrics are shown for the 597 sites where rejection succeeded and Lighthouse ran. 
The pre-interaction/post-accept differences are statistically significant for three of the four metrics: Lighthouse score, \gls{wave} errors, and contrast errors, while \gls{wave} alerts is not (two-sided Wilcoxon signed-rank tests, with all significant effects being medium-to-large in magnitude). 
For post-reject, three of the four differences are also significant: Lighthouse score, contrast errors, and \gls{wave} alerts, while the change in \gls{wave} errors is not; consistent with rejection producing only a marginal and non-systematic effect on error counts.

Across all \gls{wave} categories, 91 issue types showed a change in the number of affected sites after cookie interaction. The dominant pattern is a \emph{decrease} following both accept and reject: 72 of 91 issue types affected fewer sites post-accept, and 83 fewer sites post-reject. This indicates that cookie notices are themselves a significant source of these issues. 
A smaller number of issue types increased after interaction: 12 post-accept and 7 post-reject. The largest single increase was \emph{Spacer image missing alternative text} (+97 sites post-accept), while the largest decrease was \emph{Empty heading} (-83 sites post-accept). Post-accept changes are generally larger in magnitude than post-reject changes, consistent with acceptance triggering additional third-party content loading. 

Lighthouse flagged 34 audits whose fail counts changed after a cookie interaction. As with \gls{wave}, the majority (23 of 34) showed \emph{fewer} failures post-interaction, again reflecting the removal of cookie-notice elements from the page. However, 9 audits recorded \emph{more} failures after interaction, representing genuine accessibility regressions when the notice is dismissed. The most striking increase was \emph{Document does not have a main landmark.}, which rose by 184 sites post-accept and 118 post-reject. This suggests that on many sites, the cookie notice itself contains or wraps the page's main landmark, so dismissing it leaves the document without one. The largest decrease was \emph{Elements with visible text labels do not have matching accessible names.} (-96 sites post-accept), consistent with cookie notices frequently presenting unlabelled interactive controls.

In concrete terms, on nearly one in five sites with a cookie notice, the notice itself is the page's sole \texttt{<main>} landmark; once dismissed, screen reader users lose the primary navigational anchor that would otherwise let them skip repetitive header content and jump directly to the page content, forcing them back to sequential, item-by-item navigation through the full page. This is a failure of \gls{aria} landmark best practice and the \gls{wcag} 2.2 requirement that major page regions remain programmatically identifiable after user interaction.

\subsection{Screen reader accessibility metrics}
Each of the 2,061 scanned sites was evaluated against seven screen reader accessibility criteria (\autoref{tab:srm}). 
The highest pass rate was for \emph{Page Titled} (2,013 of 2,061, 98\,\%). The lowest was \emph{Cookie Notice Titled} (392 of 1,058, 37\,\%). This low pass rate indicates that the majority of cookie notices lack a programmatically exposed heading, meaning screen reader users encounter no structural landmark to identify the notice region and must instead rely on context to determine that a consent decision is being requested.
In practice, this means most consent regions are implemented as unnamed dialogues: on first encountering the notice, a screen reader user hears its content read aloud with no preceding role or name announcement (e.g.\ ``cookie consent dialog'') identifying what the region is for, a failure of \gls{wcag} 2.2 Success Criterion 4.1.2 (Name, Role, Value).
\begin{table}[htbp]\setlength{\tabcolsep}{3pt}\centering\footnotesize
\caption{Screen reader accessibility metric results from the automated crawl. N/A used for the 1,003 sites where no cookie notice was detected.}\label{tab:srm}
\begin{tabular}{lrrrr}
\toprule \textbf{Metric} & \textbf{Pass} & \textbf{Fail} & \textbf{N/A} & \textbf{Pass rate} \\ \midrule
  (i) Readable & 1{,}003 & 55 & 1{,}003 & 95\,\% \\
  (ii) Immediately Read & 415 & 643 & 1{,}003 & 39\,\% \\
  (iii) Keyboard Navigable & 979 & 79 & 1{,}003 & 93\,\% \\
  (iv) Link or Button Purpose & 449 & 609 & 1{,}003 & 42\,\% \\
  (v) Abbreviations Explained & 965 & 93 & 1{,}003 & 91\,\% \\
  (vi) Page Titled & 2{,}013 & 48 & 0 & 98\,\% \\
  (vii) Cookie Notice Titled & 392 & 666 & 1{,}003 & 37\,\% \\
\bottomrule\end{tabular}
\end{table}

Regarding \emph{immediately read} criteria (\autoref{tab:ird}), among the 1,003 sites where a cookie-related term was found in the \gls{nvda} transcript, the median number of words appearing before that term was 165 (mean: 455.0). The large gap between the mean and median reflects a heavily right-skewed distribution: while most sites place the notice reasonably early in the reading order, a minority place it so late that it substantially inflates the mean. 
Criterion \emph{Link or Button Purpose} had a pass rate of 42\,\%: the majority of sites presented cookie notice controls (accept and reject buttons) whose labels did not convey what the user was consenting to, using generic text (e.g., ``Accept'' or ``I agree'') without any reference to cookies, data, or tracking. This is a direct barrier to informed consent for screen reader users who cannot rely on the surrounding visual context. Criterion \emph{Cookie Notice Titled} had the lowest pass rate of (37\,\%), indicating that most notices lack a programmatically exposed heading to identify the region. Criterion \emph{Immediately Read} also fell below 50\,\% (39\,\%), meaning that on most sites a screen reader user must navigate through substantial page content before encountering the notice. In contrast, criteria \emph{Readable} (95\,\%), \emph{Keyboard Navigable} (93\,\%), and \emph{Abbreviations Explained} (91\,\%) all passed on the large majority of applicable sites, suggesting that while notices are generally perceivable and reachable, the quality of the interaction they offer falls well short of what is needed for informed consent.

\begin{table}[h]\centering\footnotesize
\caption{Distribution of word distance before the first cookie keyword.}\label{tab:ird}
\begin{tabular}{lrr}
\toprule \textbf{Words before cookie keyword} & \textbf{Sites} & \textbf{\%} \\ \midrule
  No keyword found & 55 & 5\,\% \\
  $= 0$ & 80 & 8\,\% \\
  $1$--$10$ & 285 & 27\,\% \\
  $11$--$30$ & 50 & 5\,\% \\
  $31$--$100$ & 62 & 6\,\% \\
  $> 100$ & 526 & 50\,\% \\
\midrule
  \textbf{Pass ($\leq 30$)} & 415 & 39\,\% \\
\bottomrule\end{tabular}
\end{table}
\subsection{Cross website tier comparison}
As detailed in \autoref{tab:srm-strata}, screen reader accessibility reveals a clear trade-off based on site popularity. Top-tier sites excel in interaction fundamentals like keyboard navigability and descriptive button labelling, but their complex layouts bury notices deeper in the reading order. Conversely, Bottom-tier sites announce notices earlier and use plainer language (expanding abbreviations more reliably), but fail critically at interaction design---only a quarter provide meaningful button labels, severely restricting informed choice. Foundational metrics like basic page titling remain universally high, whereas structural headings (\emph{Cookie Notice Titled}) remain uniformly deficient across all tiers. Full analysis is provided in \autoref{AppD}.

\begin{table}[b]\centering\footnotesize
\caption{Screen reader pass rates (\%) by Tranco ranking.}\label{tab:srm-strata}
\begin{tabular}{lrrr}
\toprule \textbf{Metric} & \textbf{Top} & \textbf{Middle} & \textbf{Bottom} \\ \midrule
  (i) Readable & 94\,\% & 94\,\% & 97\,\% \\
  (ii) Immediately Read & 36\,\% & 39\,\% & 44\,\% \\
  (iii) Keyboard Navigable & 96\,\% & 92\,\% & 88\,\% \\
  (iv) Link or Button Purpose & 51\,\% & 43\,\% & 27\,\% \\
  (v) Abbreviations Explained & 88\,\% & 91\,\% & 97\,\% \\
  (vi) Page Titled & 98\,\% & 98\,\% & 96\,\% \\
  (vii) Cookie Notice Titled & 33\,\% & 40\,\% & 39\,\% \\
\bottomrule\end{tabular}
\end{table}
In summary, our differential analysis exposes cookie notices as structural risks rather than only visual overlays: notices create unlabelled, ARIA-heavy widgets and keyboard traps, and dismissing them frequently strips critical page architecture, most notably the underlying document's \texttt{<main>} landmark. We return to the implications of this pattern for \gls{blv} users in \S\ref{sec:discussion}.

\section{Comparative privacy and accessibility results}
Addressing RQ3, our findings reveal a systematic relationship between \gls{ukgdpr} and \gls{pecr} compliance and screen reader accessibility: the notice designs that fail legal requirements are disproportionately those that also create the greatest barriers for \gls{blv} users, including with the weakest screen reader accessibility. The privacy harms that follow from rejection are most acute on sites where \gls{blv} users are least able to navigate the notice in the first place. Even notices that provide a direct reject path exhibit structural deficiencies that impede informed consent for \gls{blv} users. The analyses below examine the dimensions of this intersection.

Analysis across Tranco tiers reveals that site popularity dictates the type of harm a BLV user encounters, rather than its severity. High-profile (top-tier) sites invest in complex \glspl{cmp} that provide descriptive button labels, but these bloated widgets introduce severe structural regressions---most notably destroying the page's main landmark upon dismissal. Conversely, less prominent (bottom-tier) sites rely on simpler, plainer text but fail entirely at basic interactive labelling, leaving users to guess the purpose of unlabelled controls. Most critically, the middle-tier emerges as a neglected tracking hotspot: these sites deploy the most aggressive third-party tracking ecosystems both before and after consent, while simultaneously exhibiting the most structurally destructive DOM footprints. For BLV users, Top sites break their navigational architecture, Bottom sites deny them informed choice, and Middle-tier sites heavily track them while doing both. 

\subsection{UK GDPR/PECR compliance and accessibility}
Here, a notice is classified as \emph{compliant} if it offers a direct rejection path 
and the accept option is not visually emphasised relative to the reject option~\cite{ICOAcceptOrPay,UKGDPR}. 
\begin{table}[b]\setlength{\tabcolsep}{3pt}\centering\footnotesize
\caption{Screen reader pass rates by \gls{ukgdpr}/\gls{pecr} compliance status.}\label{tab:compliance_vs_srm}
\begin{tabular}{lrr}
\toprule \textbf{Metric} & \textbf{Compliant} & \textbf{Non-compliant} \\ \midrule
  (i) Readable & 96\,\% & 94\,\% \\
  (ii) Immediately Read & 39\,\% & 40\,\% \\
  (iii) Keyboard Navigable & 98\,\% & 88\,\% \\
  (iv) Link or Button Purpose & 54\,\% & 31\,\% \\
  (v) Abbreviations Explained & 92\,\% & 91\,\% \\
  (vi) Page Titled & 99\,\% & 99\,\% \\
  (vii) Cookie Notice Titled & 41\,\% & 33\,\%\\
\bottomrule\end{tabular}
\end{table}
\paragraph{Screen reader metrics.}
As shown in \autoref{tab:compliance_vs_srm}, legal compliance correlates strongly with criteria tied directly to the consent interaction: compliant notices are notably more keyboard-navigable, descriptively labelled, and structurally titled. This suggests regulatory pressure for symmetric choices encourages more deliberate, findable UI engineering. In contrast, broader site-level accessibility practices---such as general readability, abbreviation expansion, and DOM placement (\emph{Immediately Read})---remain virtually identical across both groups. Even legally compliant notices frequently fail to provide structural headings or informative button labels. Therefore, while regulatory compliance improves interaction fundamentals, it falls short of guaranteeing a fully accessible screen reader experience.

\paragraph{Automated accessibility metrics.}
Non-compliant notices are associated with worse baseline accessibility (\autoref{tab:complexity_tradeoff}), exhibiting more WAVE errors, contrast issues, and ARIA elements. While this partly reflects the generally poorer design of lower-ranked sites, non-compliant interfaces also leave a vastly larger structural footprint upon dismissal—driven by the complex preference panels common in \emph{Accept or Settings} flows. Despite this structural bloat, dismissing either notice type yields only marginal, equivalent changes (a small Lighthouse score decline and a small increase in WAVE errors, alongside the expected drop in ARIA element count as the notice's markup is removed) to overall WAVE and Lighthouse scores. For BLV users, the practical implication is stark: neither compliance group treats accessibility as a priority alongside legal adherence. Severe navigational hazards---such as missing main landmarks and unnamed dialogues---plague both groups equally, proving that legal compliance status cannot reliably predict the accessibility of the underlying page.
\begin{table*}[t]\setlength{\tabcolsep}{3pt}\centering\footnotesize
\caption{\gls{wave} metrics by \gls{ukgdpr}/\gls{pecr} compliance status (sites with a cookie notice).}\label{tab:complexity_tradeoff}
\begin{threeparttable}
\begin{tabular}{lrr}
\toprule \textbf{Metric} & \textbf{Compliant} & \textbf{Non-compliant} \\ \midrule
  Sites$^{\dag}$ & 509 & 511 \\
  Avg.\ WAVE errors (pre) & 17.2 & 20.1 \\
  Avg.\ WAVE ARIA elements (pre) & 556.2 & 636.4 \\
  Avg.\ WAVE contrast errors (pre) & 18.5 & 36.0 \\
  \addlinespace[3pt]
  Avg.\ $\Delta$ WAVE errors (post-accept) & +0.6 & +0.5 \\
  Avg.\ $\Delta$ WAVE ARIA (post-accept) & -140.5 & -358.7 \\
  Avg.\ $\Delta$ Lighthouse score (post-accept) & -1.0 & -0.8\\
\bottomrule\end{tabular}
\begin{tablenotes}\footnotesize
  \item[$\dag$] 509/511 sites of the 528/530 compliant/non-compliant totals had a recorded post-accept WAVE error score, needed to compute this table's $\Delta$ rows.
\end{tablenotes}
\end{threeparttable}
\end{table*}

\paragraph{Shared accessibility and privacy risks.}
Of the 2,061 reachable sites, 1,003 (49\,\%) displayed no cookie notice, a violation for any site serving UK users. A subset of these may be compliant under the \gls{duaa} analytics exemption if they use only exempt categories and provide notice and opt-out through other means; this analysis cannot make that distinction. This figure is consistent with prior work (e.g., ~\cite{kampanos2021accept}). 
As established in Tables \ref{tab:tracker-request-summary} and \ref{tab:accessibility_with_cookie_notices}, sites lacking a cookie notice exhibit both lower baseline accessibility and higher unconsented tracking rates. For BLV users, this co-occurrence is a compounding disadvantage: they face simultaneous navigation barriers and data collection, with neither addressable since no consent mechanism exists to interact with. This suggests a shared underlying cause, the resources and compliance awareness required to implement a cookie notice overlap substantially with those required to invest in accessibility. 

\paragraph{Accept or pay privacy notice model.}
Twenty-five sites (2\,\% of those with a notice) implement the \emph{accept or pay} model, conditioning access on either accepting non-essential cookies or paying a subscription fee. From a privacy perspective, consent bundled with service access may not be freely given within the meaning of \gls{ukgdpr}. According to the ICO, this model may fail the freely-given standard~\cite{UKGDPR,ICOAcceptOrPay}. The model also creates a structural accessibility trap for \gls{blv} users; a user who cannot identify or interact with the notice has no accessible route to either available response. 
As detailed in \autoref{tab:pay_srm}, accept-or-pay sites perform substantially worse on consent-critical criteria. Severe drops in \emph{Link or Button Purpose}, \emph{Readable}, and \emph{Abbreviations Explained} indicate that these notices rely on uninformative labels and complex legal jargon, making them frequently incomprehensible or entirely inaudible to screen readers. For BLV users, this model uniquely compounds compliance issues with concrete accessibility barriers: because the formally available choices (payment or tracking) are practically inaccessible, users are left with no actionable choice.
\begin{table}[t]\setlength{\tabcolsep}{3pt}\centering\footnotesize
\caption{Screen reader pass rates: accept-or-pay sites (25 sites, 2 \% of those with a notice) vs all sites with a notice.}\label{tab:pay_srm}
\begin{tabular}{lrr}
\toprule \textbf{Metric} & \textbf{Accept-or-pay} & \textbf{All notices} \\ \midrule
  (i) Readable & 84\,\% & 95\,\% \\
  (ii) Immediately Read & 36\,\% & 39\,\% \\
  (iii) Keyboard Navigable & 88\,\% & 93\,\% \\
  (iv) Link or Button Purpose & 8\,\% & 42\,\% \\
  (v) Abbreviations Explained & 72\,\% & 91\,\% \\
  (vi) Page Titled & 100\,\% & 99\,\% \\
  (vii) Cookie Notice Titled & 28\,\% & 37\,\% \\
\bottomrule\end{tabular}
\end{table}

\paragraph{Screen reader accessibility and post-reject tracker exposure.}
\begin{table*}[t]\centering\footnotesize
\caption{Median post-reject tracker requests by screen reader pass/fail status.}\label{tab:srm_tracker_mwu}
\begin{threeparttable}
\setlength{\tabcolsep}{2pt}
\begin{tabular}{lrrrrrr}
\toprule \textbf{Metric} & \textbf{Pass $n$} & \textbf{\makecell{Pass \\ med.}} & \textbf{Fail $n$} & \textbf{\makecell{Fail \\ med.}} & $p$ & $r$ \\ \midrule
  (i) Readable & 581 & 17 & 21 & 42 & $0.040$ & $0.08$ \\
  (ii) Immediately Read & 272 & 16 & 330 & 22 & $0.015$ & $0.10$ \\
  (iv) Link or Button Purpose & 326 & 16 & 276 & 24 & $<\!0.001$ & $0.17$ \\
  (v) Abbreviations Explained & 553 & 16 & 49 & 28 & $0.005$ & $0.12$ \\
  (vii) Cookie Notice Titled & 235 & 16 & 367 & 19 & $0.632$ & $0.02$ \\
\bottomrule\end{tabular}
\begin{tablenotes}\footnotesize
  \item Two-sided Mann-Whitney U test; $r$ = rank-biserial correlation (effect size).
\end{tablenotes}
\end{threeparttable}
\end{table*}
To test if screen reader accessibility failures compound privacy harms, we compare post-reject tracker counts across pass and fail groups for each criterion (\autoref{tab:srm_tracker_mwu}). Four of five evaluated criteria show significant differences: sites failing \emph{Link or Button Purpose}, \emph{Immediately Read}, and \emph{Abbreviations Explained} exhibit substantially higher tracker medians (24, 22, and 28) than passing sites (16). Only \emph{Cookie Notice Titled} shows no significant effect. Overall, sites where BLV users face the greatest interaction barriers provide the least privacy protection post-rejection, demonstrating compounding harms.

\begin{table*}[t]\centering\footnotesize
\caption{Screen reader pass rates by notice emphasis.}\label{tab:emph_srm}
\begin{threeparttable}
\setlength{\tabcolsep}{1.5pt}
\begin{tabular}{lrrrrrr}
\toprule \textbf{Metric} & \textbf{Acc.\ } & \textbf{Acc.\ } & \textbf{Equal} & \textbf{Equal } & $p$ & $V$ \\
&\textbf{$n$}&\textbf{\% pass}&\textbf{$n$}&\textbf{\%  pass}&&\\\midrule
  (i) Readable & 353 & 95\,\% & 649 & 94\,\% & $0.521$ & $0.02$ \\
  (ii) Immediately Read & 353 & 46\,\% & 649 & 35\,\% & $<\!0.001$ & $0.11$ \\
  (iii) Keyboard Navigable & 353 & 90\,\% & 649 & 97\,\% & $<\!0.001$ & $0.14$ \\
  (iv) Link or Button Purpose & 353 & 33\,\% & 649 & 49\,\% & $<\!0.001$ & $0.16$ \\
  (v) Abbr. Explained & 353 & 90\,\% & 649 & 91\,\% & $0.803$ & $0.01$ \\
  (vi) Page Titled & 353 & 99\,\% & 649 & 100\,\% & $0.741$ & $0.01$ \\
  (vii) Cookie Notice Titled & 353 & 33\,\% & 649 & 42\,\% & $0.003$ & $0.09$ \\
\bottomrule\end{tabular}
\begin{tablenotes}\footnotesize
  \item Chi-squared test; $V$ = Cram\'{e}r's $V$ (effect size).
\end{tablenotes}
\end{threeparttable}
\end{table*}
\subsection{Notice features and accessibility}
Cookie notice visual presentation, structure and position also have an impact on accessibility, as we discuss here. 
\paragraph{Visual emphasis and screen reader accessibility.}
Notices visually emphasising the accept option are also significantly less accessible to screen reader users (\autoref{tab:emph_srm}). Comparing accept-emphasised versus equal-weight notices reveals lower pass rates for \emph{Link or Button Purpose}, \emph{Keyboard Navigable}, and \emph{Cookie Notice Titled}. Thus, design choices acting as visual dark patterns for sighted users simultaneously create structural barriers for BLV users. The exception is \emph{Immediately Read}, where accept-emphasised notices score higher, likely because simple nudge banners typically occupy early DOM positions compared to complex full-screen dialogues.

\paragraph{Notice structural complexity and post-accept tracking.}
The structural features of a cookie notice, measured by the change in \gls{wave} \gls{aria} element count after acceptance, are negatively associated with post-accept tracker request volume.
A larger \gls{aria} delta indicates a more complex notice, consistent with the granular consent panels common in \emph{Accept, Reject, or Settings} designs.
Across 1,018 sites with both measures, there is a statistically significant weak negative correlation (Spearman's $\rho = -0.26$ ($p < 0.001$)), i.e., sites with more structurally complex notices tend to activate \emph{fewer} post-accept tracker requests.
At the individual site level, sites that invest in elaborate consent management platforms are also more likely to gate tracker activation on consent correctly, while sites with simple accept banners may load third-party tracking infrastructure regardless of notice design.
The negative correlation, therefore, does not imply that notice complexity protects privacy; it reflects, rather, how seriously a site treats regulatory compliance overall.

\paragraph{Notice position and screen reader accessibility.}
The cookie notice position can determine where the notice appears in the \gls{dom} reading order encountered by a screen reader.
\autoref{tab:position_srm} shows pass rates by position for three of the most position-sensitive criteria.
The overall chi-squared test for \emph{Immediately Read} across positions is not statistically significant, 
indicating that visual placement alone does not determine where the notice falls in the screen reader reading order.
\begin{table}[t]\setlength{\tabcolsep}{3pt}\centering\footnotesize
\caption{Screen reader pass rates by cookie notice position; O for Overlay.}\label{tab:position_srm}
\begin{tabular}{lrrrr}
\toprule \textbf{Position} & \textbf{$n$} & \textbf{\makecell[l]{\% (ii)\\  Imm. Read}} & \textbf{\makecell[l]{\% (iv)\\ Link Purpose}} & \textbf{\makecell[l]{\% (vii) \\ CN Titled}} \\ \midrule
  Bottom O & 460 & 37\,\% & 43\,\% & 30\,\% \\
  Corner O & 123 & 35\,\% & 51\,\% & 28\,\% \\
  Middle O & 357 & 45\,\% & 32\,\% & 48\,\% \\
  Top O & 100 & 34\,\% & 66\,\% & 41\,\% \\
\bottomrule\end{tabular}
\end{table}
The most notable pattern is on \emph{Link or Button Purpose}, where Top Overlay notices score substantially higher than Middle Overlay notices. Top overlays are typically simple two-button banners whose labels are more descriptive; Middle Overlay notices are more often full-screen consent dialogues with generic call-to-action text. The Middle Overlay position also has the highest \emph{Immediately Read} pass rate consistent with these notices being rendered as modal dialogues early in the \gls{dom}, while having the worst \emph{Link or Button Purpose} performance. Hence, there is a positional trade-off: the notice positions that place the consent decision early in the reading order tend to be the same ones that provide the least informative button labels. This is likely not accidental; prior work found that barrier-style notices, such as middle overlays, were ignored 3.6 times less often than banner-style notices~\cite{utzInformedConsentStudying2019}, suggesting that position choices are driven by consent-maximisation goals for sighted users rather than by information clarity. The consequence for \gls{blv} users is that the positions most effective at securing sighted-user engagement happen to pair early reading-order placement with the weakest button labelling, providing structure but not the meaning needed to act on it.

Our analysis shows that \gls{blv} users barriers are not random but systematic properties of how cookie notices are engineered. The same design decisions that reduce navigability also correlate with weaker post-interaction privacy protection. This coherence suggests that accessibility and privacy in cookie notice design are best treated as a unified engineering concern rather than two separate compliance obligations.

\section{Recommendations for Practitioners}
\label{sec:discussion}
This section examines what the above failures mean for \gls{blv} users and sets out concrete recommendations for stakeholders.

Our results show that cookie notices impose two compounding barriers on \gls{blv} users.
First, a \emph{pre-interaction barrier}: an untitled cookie notice (63\,\% fail \emph{Cookie Notice Titled}), uninformative controls (58\,\% fail \emph{Link or Button Purpose}), and placement deep in the reading order (61\,\% fail \emph{Immediately Read}) together prevent an informed decision before one is even made. Second, \emph{post-interaction barrier}: dismissing the notice removes the page's only \texttt{<main>} landmark on one in five sites, degrading navigation for the remainder of the browsing session regardless of which option was chosen. These correspond closely to dark patterns previously identified in \gls{at} contexts~\cite{lewis_inaccessible_2025}: unlabelled controls function as \emph{misdirection}, and a notice buried in the reading order functions as \emph{obstruction}, with the enforced sequential reading of a screen reader turning that obstruction into \emph{assistive amplification}~\cite{stanczyk_revealed_2026} rather than a minor inconvenience.

Based on our findings, we provide the following recommendations for the multiple stakeholders involved in this complex issue.

\paragraph{Consent management developers.} Three structural fixes would resolve the majority of the interaction-quality failures we observe. (i) The notice should expose a programmatic heading identifying it as a consent region (\emph{Cookie Notice Titled} fails on 63\,\% of sites); this is a single \gls{aria} attribute or native heading element, not a redesign. (ii) Response controls should carry descriptive accessible names (e.g.\ ``Accept all cookies'' rather than ``OK'') rather than relying on visual context alone (\emph{Link or Button Purpose} fails on 58\,\%). (iii) Developers should ensure the notice is not the page's sole \texttt{<main>} landmark, or that an equivalent landmark is restored once the notice is dismissed, since its removal otherwise strips every subsequent page in the session of its primary navigational anchor for screen reader users (affecting one in five sites with a notice). 
\\
None of these fixes requires relocating the notice visually; a notice can remain a small corner or bottom overlay for sighted users while still being announced immediately and titled correctly for \gls{at} users, so correcting the markup does not trade off against a site's existing consent-rate design goals.

\paragraph{Standards bodies.} The \gls{wcag} does not currently include success criteria specific to consent interfaces, despite cookie notices now being near-ubiquitous, transient, and interaction-critical UI elements that general criteria such as 4.1.2 (Name, Role, Value) and 2.4.6 (Headings and Labels) only partially capture. A targeted technique or success criterion, covering notice titling, immediate announcement, and preservation of page landmark structure on dismissal, would give auditors and automated tools such as \gls{wave} and Lighthouse a direct target; currently, neither tool flags most of the failures we identify (\S\ref{sec:accessibility-results}), so sites can pass automated accessibility audits while still presenting an inaccessible consent interaction. The \gls{wcag} 3.0 working draft's Guideline 2.10 (Policy and protection), which explicitly calls for protecting users' privacy and security, is a rare acknowledgement of this gap at the standards level~\cite{wcag3}, though its scope and enforceability for consent interfaces specifically remain to be determined.

\paragraph{Regulators and policymakers.} The \gls{eaa} has applied to a defined range of consumer-facing private-sector services since June 2025, which brings consent interfaces within its scope; our sample, crawled after that date, shows the accessibility failures above remain widespread, suggesting \gls{eaa} obligations have not yet translated into accessible consent design in practice. More directly, our results give regulators concrete, empirical, automated evidence for enforcement. The absence of a reject path, visual de-emphasis of the reject option, and the accept-or-pay model are all automatically detectable at scale using our methodology. 
\\
Evidently, the correlation we find between \gls{ukgdpr}/\gls{pecr} non-compliance and lower screen reader accessibility (\S\ref{sec:accessibility-results}) suggests that \textbf{privacy compliance and accessibility enforcement would reinforce rather than compete with one another if pursued jointly.} The \gls{duaa}'s new opt-out exemptions for analytics and security cookies do not relieve any of the violations we identify, since they still require an accessible opt-out mechanism; our finding that only 42\,\% of notices use descriptive button labels raises the question of whether opt-out mechanisms introduced under the \gls{duaa} will themselves be accessible to \gls{blv} users, which we leave as an open question for regulatory guidance and future work. \\
More broadly, \gls{ukgdpr}/\gls{pecr} and \gls{wcag} were developed independently and, in practice, create tension rather than a coherent joint framework. Prior work shows that \gls{wcag} and the \gls{eaa} already implicate dark pattern categories such as hidden information and auto-play~\cite{pellkvist_access_2026}, suggesting that the deceptive properties we identify in cookie notices, unlabelled controls, visually emphasised accept buttons, and notices buried in the reading order, may already fall within the scope of existing accessibility legislation, giving regulators a lever beyond \gls{ukgdpr}/\gls{pecr} enforcement alone.

\section{Conclusion} \label{sec:conclusion}
This study deployed an automated measurement pipeline, successfully evaluating over 2,000 reachable domains to measure the intersection of cookie notice privacy and accessibility for BLV users. Our findings reveal a landscape of compounding harms. While 51 \% of sites display a cookie notice, half of these violate UK GDPR and PECR by obscuring or omitting reject options, and the remaining 49 \% provide no consent mechanism at all. Accepting a notice triggers a five-fold increase in tracking, yet rejecting it remains disproportionately difficult for screen reader users. Although 95 \% of notices are technically readable, they fail critically on interaction quality: most lack structural titles, use uninformative button labels, and are buried deep in the reading order. Furthermore, dismissing these notices often degrades the underlying page's accessibility, such as by stripping the main navigational landmark.

The interfaces that fail legal privacy standards also present the most severe barriers to AT. For BLV users, this translates to a dual disadvantage: a structural inability to exercise informed consent, coupled with heightened exposure to unconsented tracking. The findings of this paper establish a quantitative baseline for cookie notice privacy and accessibility at scale. This work underscores the urgent need for policymakers, regulators, and developers to treat web privacy and accessibility as a unified engineering imperative.

\begin{acks}
Claude was used to assist with troubleshooting the code for this project, while Grammarly and Gemini Pro were used for grammar checking and writing refinement. The core coding and writing were conducted by the authors. All AI-assisted work was thoroughly reviewed, manually verified, and significantly refined by the authors prior to the final experiments and manuscript submission.
\end{acks}

\section*{Conflicts of Interest}
The authors declare that there is no conflict of interest regarding the publication of this article.

\section*{Funding Statement}
This research received no specific funding.

\begin{ethics}
This study relies exclusively on the automated analysis of publicly accessible websites. Because the research did not involve human subjects, user data, or user interactions, %
no Personally Identifiable Information (PII) was collected or stored. 
To ensure our measurement infrastructure did not negatively impact the surveyed websites or cause a denial-of-service effect, our crawler was designed to be minimally invasive: we restricted our analysis to the root landing pages, did not employ aggressive concurrent scraping against individual domains, and did not bypass technical access controls. 

To balance scientific rigour with environmental responsibility, we carefully scoped our dataset to a sample of 3,000 websites. Also, we captured and preserved all raw artifacts from these crawls—including network traffic, HTML/CSS, cookies, client-side storage, and screen reader transcripts for our local large-scale analysis. By releasing this comprehensive dataset for further analysis, we enable future research without the need for redundant crawling.%
\end{ethics}

\begin{openscience}
The source code for the analyses in this paper is available at \url{https://github.com/JamesMClarke/automated_cookie_notices}.
The full dataset collected during the crawls described in this paper is available at \url{https://osf.io/zyfku/overview}.
\end{openscience}

\bibliographystyle{plainnat}
\bibliography{references}

\begin{appendix}
\renewcommand{\thetable}{\thesection\arabic{table}}
\renewcommand{\thefigure}{\thesection\arabic{figure}}

\section{Cookie notice accessibility metrics}
\setcounter{figure}{0}\setcounter{table}{0}
\label{appendix:cookie-accessibility-metrics}

The definitions of seven metrics for cookie notice accessibility evaluation are based on \cite{ClarkeInvisible2024} and changed to fit our methodology, i.e., to make them better defined for automatic measurements.

\emph{(i) Readable.} A cookie notice was considered readable if its content was accessible to the screen reader. Operationally, this was assessed by checking for the presence of cookie-related terms within the \gls{nvda} transcript captured before any interaction with the page. The detection vocabulary mirrors BannerClick's cookie-identification word list and its multilingual extensions (German, French, Spanish, Italian, Dutch, Portuguese)~\cite{rasaii2023cookiewall,rasaii2023exploring}, covering terms such as \emph{cookie}, \emph{consent}, \emph{privacy}, \emph{gdpr}, \emph{tracking}, \emph{datenschutz}, \emph{consentement}, \emph{privacidad}, and their equivalents across six further languages. Sites where no cookie notice was detected were excluded from this metric. 

\emph{(ii) Immediately Read.} A cookie notice passes if it is the first content encountered by the screen reader after page load. We measured how many words appeared in the transcript before the first occurrence of any term from the cookie-identification vocabulary described in criterion~(i). Image-placeholder tokens produced by the screen reader were stripped before counting. A site passed if at most 30 words preceded the first such term. We set the threshold at 30 words based on experience from our earlier work~\cite{ClarkeInvisible2024}: a page title is typically 3--8 words; a single landmark announcement (e.g.\ ``banner landmark'') adds 1--2 words; and a short tagline or navigation label could contribute a further 5--15 words. The 30-word ceiling comfortably accommodates these elements while flagging notices buried deeper in the reading order.

\emph{(iii) Keyboard Navigable.} A cookie notice passes if it is navigable using a keyboard while using a screen reader. Each scan produced a list of all keyboard-focusable elements on the page prior to any interaction, collected by injecting JavaScript that queries \texttt{button}, \texttt{a[href]}, \texttt{input}, \texttt{select}, \texttt{textarea}, and any element carrying a non-negative \texttt{tabindex}, filtered to those that are enabled and have a non-zero layout footprint or a valid offset parent. Each element was additionally tagged to indicate whether it resided inside an ancestor carrying \texttt{role=``dialog''} or \texttt{aria-modal=``true''}. A site passed if at least one such in-dialogue element was present, indicating that the cookie notice could be reached via keyboard navigation and that interactive controls within it were exposed in the tab order. Where no dialogue-scoped elements were found, the metric fell back to checking whether any focusable element on the page carried an action-word label from a multilingual vocabulary covering English, German, French, Spanish, Italian, Dutch, Portuguese, and Norwegian.

\emph{(iv) Link or Button Purpose.} A cookie notice passes if the purpose of each link or button can be determined from the link or button text alone. The accept and reject controls were identified by matching their accessible label or visible text against the same multilingual action-word vocabulary used for metric~(iii). Each identified control was then checked to determine whether its label, considered in isolation, conveyed what the user would be consenting to. A label was considered sufficiently informative if it contained at least one subject term from the set \emph{cookie(s)}, \emph{data}, \emph{privacy}, \emph{tracking}, \emph{consent}, \emph{essential}, \emph{necessary},  \emph{analytics}, \emph{marketing}, \emph{advertising}, \emph{functional}, \emph{optional}, or \emph{performance}. A site passed if at least one action control carried a descriptive label such as ``Accept all cookies'' or ``Decline optional cookies''; labels such as ``Accept'', ``I agree'', or ``OK'' without a subject term caused the site to fail.

\emph{(v) Abbreviations Explained.} A cookie notice passes if all abbreviations are explained. The full text of the screen reader transcript was scanned for a predefined list of regulatory and technical abbreviations commonly found in cookie notices: \gls{gdpr}, \gls{ccpa}, \gls{pecr}, IAB, TCF, \gls{cmp}, \gls{dpa}, \gls{ico}, \gls{cpra}, LGPD, and DSA. Whole-word matching was used to avoid false positives. Where an abbreviation was detected, the transcript was first checked for the presence of its expanded form in the body text; if no expansion was found, the link list was searched for any link whose label contained the abbreviation, on the basis that a labelled link to an explanatory resource constitutes an adequate reference. A site passed if every detected abbreviation was either expanded in-line or referenced by a labelled link, or if no listed abbreviations appeared.

\emph{(vi) Page Titled.} A page passes if it has a title that describes its topic or purpose. This metric was evaluated for all scanned sites regardless of cookie notice presence. Operationally, a page was considered titled if the \texttt{statistics.pagetitle} field of the \gls{wave} accessibility report was non-empty, indicating that a meaningful \texttt{<title>} element was present in the document.

\emph{(vii) Cookie Notice Titled.} A cookie notice passes if it has a title or heading readable by the screen reader. The \gls{nvda} transcript exposes a \texttt{headings} array containing all heading elements found in the accessibility tree. A site passed if at least one heading contained any term from the cookie-identification vocabulary described in criterion~(i), indicating that the notice region carried a programmatically exposed heading readable by the screen reader.
\section{Unavailable websites} \label{sec:unavailable-websites}
\setcounter{figure}{0}\setcounter{table}{0}
Of the three thousand sites scanned, 939 failed to load. The most common failure was \emph{DNS not resolved} (498 sites), which typically indicates infrastructure or CDN domains (e.g.\ \texttt{akamai.net}, \texttt{akadns.net}) that do not serve end-user web pages. All unavailable sites were excluded from the cookie notice and accessibility analysis. \autoref{tab:errors} provides an overview of error categories encountered when loading sites in Chrome, along with their distribution across the three popularity levels. Each error category is described in more detail below.
\begin{table}[H]\setlength{\tabcolsep}{3pt}
    \centering
    \footnotesize
    \caption{Categories of errors encountered when loading sites in Chrome.}
    \begin{tabular}{lrrrr}
        \toprule
        \textbf{Error Category} & \textbf{Top} & \textbf{Middle} & \textbf{Bottom} & \textbf{Total} \\ \midrule
        DNS not resolved & 219 & 158 & 121 & 498 \\
        Timed out & 38 & 33 & 60 & 131 \\
        HTTP 403 Forbidden & 18 & 38 & 43 & 99 \\
        TLS error & 26 & 22 & 45 & 93 \\
        Connection failed & 9 & 14 & 27 & 50 \\
        HTTP 404 Not Found & 20 & 9 & 10 & 39 \\
        Other 5xx server error & 4 & 5 & 6 & 15 \\
        Other 4xx client error & 7 & 3 & 3 & 13 \\
        Other error & 0 & 1 & 0 & 1 \\
        \bottomrule
    \end{tabular}
    \label{tab:errors}
\end{table}

\begin{description}
  \item[\emph{DNS not resolved.}] The domain name could not be resolved to an IP address, meaning the site has no public DNS record. This is the largest category and predominantly reflects infrastructure domains (CDN endpoints, ad-tech hostnames, and API subdomains) that are not intended to serve user-facing web pages.
  \item[\emph{Timed out.}] The browser failed to receive a complete response within the 30-second navigation timeout. This covers genuinely slow or unreachable servers, as well as sites that cause the browser renderer to exhaust memory (the ``Aw, Snap!'' out-of-memory state), rendering the tab unresponsive.
  \item[\emph{HTTP 403 Forbidden.}] The server responded but refused to serve the page. This typically indicates bot-detection or IP-blocking measures that prevent automated browsers from accessing the site.
  \item[\emph{TLS error.}] The HTTPS handshake could not be completed. Common causes include expired certificates, certificates issued for a different domain (SNI mismatch), untrusted certificate authorities, and servers that only support deprecated cypher suites.
  \item[\emph{Connection failed.}] A TCP connection could not be established or was terminated before the response was received. This includes servers actively refusing connections, mid-flight connection resets, empty responses, and redirect loops.
\end{description}
The remaining categories (HTTP 404, other 5xx/4xx responses, and unclassified errors; 68 sites combined, 7\,\% of failures) are standard HTTP-level failures with no crawl-specific cause to report.
\section{Detailed privacy results}
\setcounter{figure}{0}\setcounter{table}{0}
\begin{table}[H]
  \centering
  \scriptsize
  \setlength{\tabcolsep}{3pt}
  \caption{Most common accept (a) and reject (b) button labels.}
  \label{tab:labels}
\begin{minipage}[t]{0.54\textwidth}\centering
\textbf{(a) Accept.}\\[4pt]
\begin{tabular}{lrr}
\toprule \textbf{Label} & \textbf{$n$} & \textbf{\% of sites} \\ \midrule
  Accept all & 211 & 19.9\,\% \\
  Accept & 110 & 10.4\,\% \\
  Accept all cookies & 105 & 9.9\,\% \\
  Consent & 105 & 9.9\,\% \\
  I accept & 30 & 2.8\,\% \\
  Accept cookies & 20 & 1.9\,\% \\
  Agree & 15 & 1.4\,\% \\
  Allow all & 14 & 1.3\,\% \\
  Confirm my choices & 12 & 1.1\,\% \\
  Accept additional cookies & 7 & 0.7\,\% \\
  I agree & 6 & 0.6\,\% \\
  Allow all cookies & 5 & 0.5\,\% \\
  Agree and close & 5 & 0.5\,\% \\
  Alle akzeptieren (German for Accept all) & 4 & 0.4\,\% \\
  Aceptar (Spanish for Accept) & 4 & 0.4\,\% \\
  Tümünü kabul et (Turkish for Accept all) & 4 & 0.4\,\% \\
  Akzeptieren (German for Accept) & 4 & 0.4\,\% \\
  Accept payments & 3 & 0.3\,\% \\
  Accepter et fermer (French for Accept and close) & 3 & 0.3\,\% \\
  Allow & 3 & 0.3\,\% \\
  \textit{Other} & 147 & 13.9\,\% \\
\midrule
  \textbf{Total identified} & 817 & 77.2\,\% \\
\bottomrule\end{tabular}
\end{minipage}\hfill
\begin{minipage}[t]{0.44\textwidth}\centering
\textbf{(b) Reject.}\\[4pt]
\begin{tabular}{lrr}
\toprule \textbf{Label} & \textbf{$n$} & \textbf{\% of sites} \\ \midrule
  Reject all & 222 & 21.0\,\% \\
  Reject & 45 & 4.3\,\% \\
  Decline & 39 & 3.7\,\% \\
  Necessary & 39 & 3.7\,\% \\
  Reject all cookies & 29 & 2.7\,\% \\
  Reject non-essential & 24 & 2.3\,\% \\
  Decline all & 23 & 2.2\,\% \\
  Reject additional cookies & 8 & 0.8\,\% \\
  Deny & 8 & 0.8\,\% \\
  Reject optional cookies & 7 & 0.7\,\% \\
  Disagree & 6 & 0.6\,\% \\
  Essential & 5 & 0.5\,\% \\
  Reject non-essential cookies & 5 & 0.5\,\% \\
  Ablehnen (German for Refuse)& 5 & 0.5\,\% \\
  Decline optional cookies & 4 & 0.4\,\% \\
  Decline cookies & 4 & 0.4\,\% \\
  Decline non-essential cookies & 4 & 0.4\,\% \\
  Essential cookies only & 4 & 0.4\,\% \\
  Weigeren (Dutch for Refuse) & 4 & 0.4\,\% \\
  Accept only essential cookies & 3 & 0.3\,\% \\
  \textit{Other} & 112 & 10.6\,\% \\
\midrule
  \textbf{Total identified} & 600 & 56.7\,\% \\
\bottomrule\end{tabular}
\end{minipage}
\end{table}

\begin{figure}[H]
  \centering
  \includegraphics[width=0.8\linewidth,keepaspectratio]{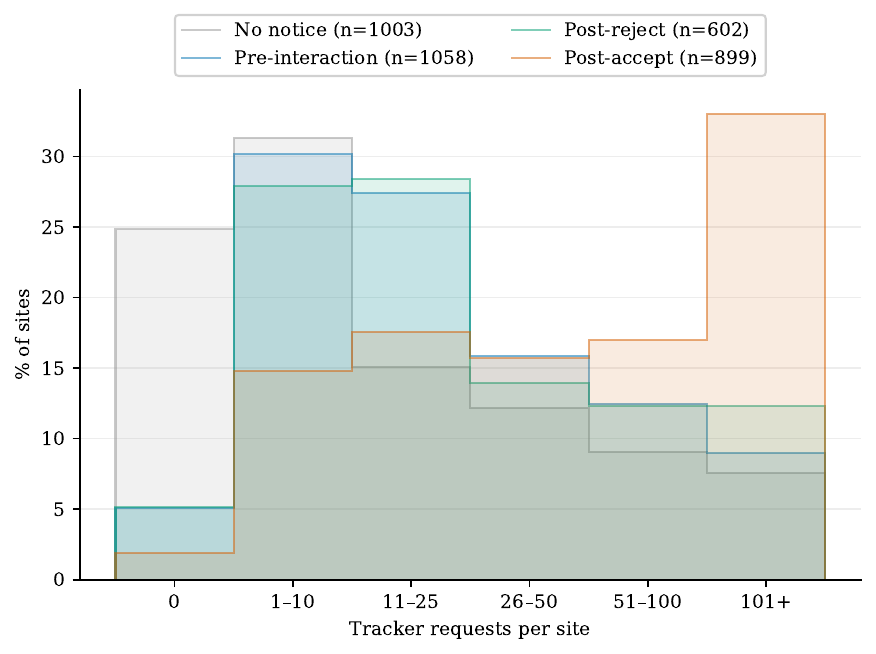}
  \caption{Distribution of sites by tracker request count range across phases.}
  \label{fig:tracker_dist_grouped}
\end{figure}
\begin{figure}[H]
  \centering
  \includegraphics[width=0.8\linewidth,keepaspectratio]{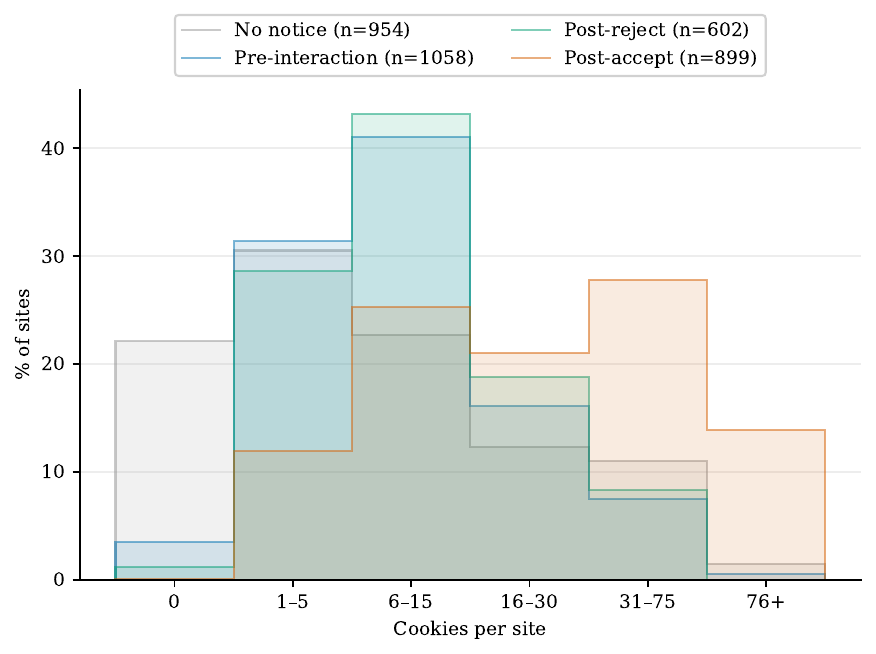}
  \caption{Distribution of sites by cookie count range across phases.}
  \label{fig:cookie_dist_grouped}
\end{figure}

\section{Detailed accessibility results}
\setcounter{figure}{0}\setcounter{table}{0}

\begin{table}[H]\setlength{\tabcolsep}{3pt}\centering\footnotesize
\caption{Mean accessibility metrics by Tranco ranking and interaction phase.}\label{tab:a11y-strata}
\begin{threeparttable}
\begin{tabularx}{\textwidth}{l XXX XXX XXX}%
\toprule
  & \multicolumn{3}{c}{Pre-interaction} & \multicolumn{3}{c}{Post-accept} & \multicolumn{3}{c}{Post-reject} \\
  \cmidrule(lr){2-4}\cmidrule(lr){5-7}\cmidrule(lr){8-10}
  \textbf{Metric} & \textbf{Top} & \textbf{Middle} & \textbf{Bottom} & \textbf{Top} & \textbf{Middle} & \textbf{Bottom} & \textbf{Top} & \textbf{Middle} & \textbf{Bottom} \\ \midrule
  Reachable sites & 659 & 717 & 685 & \multicolumn{3}{c}{---} & \multicolumn{3}{c}{---} \\
  Cookie notices detected & 63\,\% & 55\,\% & 36\,\% & \multicolumn{3}{c}{---} & \multicolumn{3}{c}{---} \\
  \addlinespace[3pt]
  LH score & 89.7 & 88.5 & 86.7 & 89.2 & 88.3 & 86.5 & 89.8 & 90.0 & 87.9 \\
  WAVE errors & 15.4 & 17.7 & 25.7 & 14.3 & 15.9 & 18.5 & 13.5 & 10.0 & 22.2 \\
  WAVE contrast errors & 23.4 & 28.5 & 32.7 & 21.8 & 26.4 & 28.1 & 21.6 & 18.1 & 17.8 \\
  WAVE alerts & 74.9 & 96.2 & 108.0 & 67.5 & 76.5 & 73.6 & 63.3 & 73.9 & 68.4 \\
\bottomrule\end{tabularx}
\begin{tablenotes}\footnotesize
  \item LH\,=\,Lighthouse score (0--100); higher is better. \gls{wave} metrics: lower is better.
\end{tablenotes}
\end{threeparttable}
\end{table}

\section{Detailed results across website tiers}
\setcounter{figure}{0}\setcounter{table}{0}
\label{AppD}
\subsection{Privacy results}
The three crawls cover distinct tiers of the Tranco ranking, referred to here as the \emph{Top} (ranks 1--1{,}000), \emph{Middle} (a random sample from ranks 1{,}001--10{,}000), and \emph{Bottom} (a random sample from beyond rank 10{,}000). Comparing them (\autoref{fig:tier_metrics} ) reveals how cookie notice prevalence, compliance, and tracking behaviour vary with site popularity.

Cookie notice detection falls steadily with Tranco rank: 63\,\% of reachable top-tier sites showed a notice, dropping to 55\,\% in the middle tier and 36\,\% in the bottom tier. This likely reflects the concentration of \gls{ukgdpr}/\gls{pecr}-facing UK and multinational services at the top of the ranking. %
Automated rejection success also declines with rank (79\,\% to 52\,\% to 36\,\%), suggesting that lower-ranked sites use less standardised notice designs that are harder for automation to navigate. The proportion of purely informational notices (offering no choice at all) rises from 4\,\% in the top tier to 10\,\% in the bottom tier. Another reason for the decline in automated rejection is the increase in notices in different languages, meaning the rejection process, which relies on keyword matching, may not identify the correct button to click.
Post-accept cookie load and tracker rate peak in the \emph{Middle} tier (avg.\ 61.2 cookies post-accept, 60.6\,\% tracker rate), exceeding both the top tier (49.3 cookies, 48.1\,\%) and the bottom tier (35.4 cookies, 52.4\,\%). Middle-tier sites, therefore, appear to deploy the most aggressive post-consent tracking, despite being less studied than the top tier.

\begin{table}[H]\setlength{\tabcolsep}{3pt}\centering\footnotesize
\caption{Per-site post-accept tracker request distribution by tier.}\label{tab:tracker_dist}
\begin{tabular}{lrrr}
\toprule \textbf{Metric} & \textbf{Top} & \textbf{Middle} & \textbf{Bottom} \\ \midrule
  No. of sites (post-accept phase) & 374 & 330 & 195 \\\midrule
  Mean trackers & 183.6 & 276.0 & 188.3 \\
  Median trackers & 42 & 62 & 42 \\
  90th percentile & 506 & 859 & 370 \\
  95th percentile & 997 & 1{,}511 & 1{,}059 \\
  Maximum & 2{,}970 & 3{,}719 & 4{,}790 \\
  \addlinespace[3pt]
  Top-10 sites' share of total & 28.1\,\% & 30.7\,\% & 55.1\,\% \\
  Mean excl.\ top 10\,\% of sites & 76.5 & 111.7 & 63.9 \\
\bottomrule\end{tabular}
\end{table}

To test whether the elevated middle-tier mean is an artefact of a small number of high-volume outlier domains, \autoref{tab:tracker_dist} disaggregates the per-site post-accept tracker request distribution across tiers. The middle tier has the highest median tracker count per site (62 requests, vs\ 42 for the top tier and 42 for the bottom tier) and the highest trimmed mean after excluding the top 10\,\% of sites by tracker count (111.7 vs\ 76.5 and 63.9). The elevated mean therefore reflects a genuine distributional shift: the typical middle-tier site loads more post-accept tracker requests than the typical site in either of the other tiers, not merely the influence of a handful of extreme outliers. The top-10 site concentration---the share of total post-accept tracker requests attributable to the ten highest-volume sites---is broadly similar for the top and middle tiers (28.1\,\% and 30.7\,\% respectively), whereas the bottom tier is considerably more concentrated (55.1\,\%), indicating that the bottom-tier mean is the least representative figure of the three.

Before any interaction, \emph{Bottom}-ranking sites carry fewer cookies per site (10.1) than Top and Middle sites (13.6 each), consistent with lower ad-tech integration on less prominent sites. The pre-interaction tracker rate tells a different story, however: \emph{Middle}-ranking sites have the highest rate (26.7\,\%) compared with 19.7\,\% for Top and 20.5\,\% for Bottom, suggesting they carry a disproportionate tracking load even before consent is sought. Post-rejection, cookie counts remain similar across tiers (Top: 13.3, Middle: 11.9, Bottom: 12.9), with no tier showing a markedly elevated post-reject cookie load; rejection appears comparably (in)effective at limiting cookie-setting regardless of site popularity. The post-reject tracker rate follows a different ordering (Middle: 25.5\,\%; Top: 21.1\,\%; Bottom: 20.9\,\%), indicating that \emph{Middle}-ranking sites continue to run the highest tracker activity even after a user has declined consent.

\subsection{Accessibility results}
This section explores how accessibility metrics differ across the three tiers of the Tranco ranking (top, middle, and bottom). %
We consider three pairwise comparisons: pre-interaction vs\ post-accept, pre-interaction vs\ post-reject, and post-accept vs\ post-reject.

\subsubsection{\gls{wave} and Lighthouse metrics}
We see a decline in baseline accessibility metrics with lower rank: the top ranking had the highest pre-interaction Lighthouse scores (avg.\ 89.7) and the fewest \gls{wave} errors (avg.\ 15.4), while the bottom ranking had the lowest Lighthouse scores (avg.\ 86.7) and the most \gls{wave} errors (avg.\ 25.7). This indicates that less popular sites tend to have poorer accessibility even before considering cookie notice interactions. With this trend holding across all three interaction phases, it suggests that the accessibility challenges faced by less popular sites are not solely due to cookie notices but may reflect broader issues in their design and development practices. \autoref{tab:a11y-strata} summarises these findings.

Post-accept, Lighthouse scores declined on substantially more sites than they improved across all three strata (Top: 48\,\% declined vs\ 16\,\% improved; Middle: 53\,\% declined vs\ 14\,\% improved; Bottom: 49\,\% declined vs\ 16\,\% improved), 
suggesting that accepting the cookie notice introduces accessibility regressions regardless of site popularity. The pattern for \gls{wave} errors is more nuanced: top-ranking sites were more likely to see errors worsen than improve (21\,\% vs\ 6\,\%), whereas this reverses in the bottom ranking, where more sites improved (15\,\%) than worsened (7\,\%). This may reflect that lower-ranked sites tend to use simpler notice designs whose removal has a net positive effect on \gls{wave} counts, while top-ranked sites deploy more complex notices that introduce errors when dismissed.

The rejection phase shows a qualitatively similar Lighthouse pattern: scores declined on more sites than they improved in all three tiers post-reject (Top: 43\,\% declined vs\ 16\,\% improved; Middle: 54\,\% vs\ 12\,\%; Bottom: 51\,\% vs\ 12\,\%). \gls{wave} error changes are substantially more balanced after rejection, with roughly equal numbers of sites improving and worsening across all tiers (Top: 8\,\% worsened vs\ 7\,\% improved; Middle: 6\,\% vs\ 5\,\%; Bottom: 6\,\% vs\ 9\,\%). Comparing post-accept directly to post-reject, rejection produces fewer \gls{wave} errors than acceptance on substantially more sites (Top: 20\,\%, Middle: 17\,\%, Bottom: 8\,\%), indicating that acceptance triggers additional error-causing content loading beyond the notice removal effect shared by both paths. %

\begin{table}[H]\setlength{\tabcolsep}{3pt}\centering\footnotesize
\caption{Top 10 \gls{wave} error, contrast, and alert issue changes post-accept by Tranco ranking.}\label{tab:wave-strata}
\begin{threeparttable}
\begin{tabular}{llrrr}
\toprule \textbf{WAVE issue ($\Delta$ post-accept)} & \textbf{Cat.} & \textbf{Top} & \textbf{Middle} & \textbf{Bottom} \\ \midrule
  Very low contrast                     & Co & -35 & -65 & -53 \\
  Redundant link                        & Al & -32 & -55 & -52 \\
  Empty heading                         & Er & -22 & -52 & -45 \\
  Suspicious link text                  & Al & -17 & -44 & -33 \\
  \makecell[l]{Spacer image missing \\alternative text} & Er & +48 & +34 & +3 \\
  Missing form label                    & Er & -15 & -41 & -27 \\
  Redundant title text                  & Al & -24 & -24 & -34 \\
  Skipped heading level                 & Al & -22 & -34 & -22 \\
  Empty button                          & Er & -14 & -28 & -35 \\
  Noscript element                      & Al & -10 & -31 & -33 \\
\bottomrule\end{tabular}
\begin{tablenotes}\footnotesize
  \item Cat.:\ Al\,=\,Alert; Co\,=\,Contrast; Er\,=\,Error.
\end{tablenotes}
\end{threeparttable}
\end{table}

The overall pattern across \gls{wave} issues (\autoref{tab:wave-strata}) is broadly consistent: all top-ten issue types decrease post-accept in all three tiers, driven by cookie notice elements being removed from the page. %
For Lighthouse audits, the largest regression across all tiers is \emph{Document does not have a main landmark} (\emph{Top}: $+57$, \emph{Middle}: $+69$, \emph{Bottom}: $+33$; these tier figures restrict the post-accept/post-reject phases to sites with a confirmed successful interaction, so they sum to less than the $+184$/$+118$ reported in \S\ref{sec:accessibility-results}, which pools all post-interaction scans regardless of confirmed success), indicating that cookie notices frequently serve as the page's sole landmark element; their removal leaves the document without navigational structure for screen reader users. The largest improvements are concentrated in label and colour contrast issues: \emph{Elements with visible text labels do not have matching accessible names} decreased by $-40$, $-57$, and $-32$ across the three tiers respectively, while colour contrast failures fell by $-39$, $-70$, and $-45$, suggesting these deficiencies originate within the notices themselves. A \emph{Top}-tier-specific regression is \emph{`[aria-hidden="true"]` elements contain focusable descendants} ($+33$ vs\ $+16$ and $+6$ for \emph{Middle} and \emph{Bottom}), consistent with the complex consent management platforms deployed by high-profile sites introducing problematic \gls{aria} containment patterns.

\section{Stability of results}\label{sec:discussion-stability}
\setcounter{figure}{0}\setcounter{table}{0}
To validate the reliability of our measurements, we repeated 25 independent crawls of \url{https://google.com}.
Google was chosen as a representative high-traffic site with a clear and detectable cookie notice.

\subsection{Privacy tests}

The cookie notices were successfully accepted in all 25 crawls and rejected in all 25 crawls, and their attributes, position, control type, emphasised option, and the availability of reject and settings options, were identical across every crawl, confirming that the page presents a consistent notice on each visit.

\begin{figure}[H]
  \centering
  \includegraphics[width=0.7\linewidth,keepaspectratio]{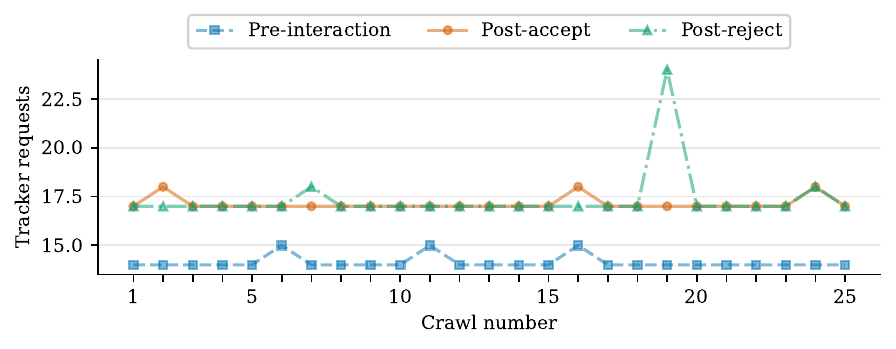}
  \caption{Tracker requests per crawl across 25 repeated crawls.}
  \label{fig:stability_trackers}
\end{figure}

Tracker request counts were also highly stable across crawls (\autoref{fig:stability_trackers}). Pre-interaction, a mean of 14.1 tracker requests were observed per crawl (range: 14--15). Post-accept, this rose to 17.1 (range: 17--18), and post-reject to 17.4 (range: 17--24). The increase in tracker requests following cookie notice dismissal, regardless of whether the notice was accepted or rejected, was consistent across all 25 crawls.

\begin{table}[H]\setlength{\tabcolsep}{3pt}\centering\footnotesize
\caption{Stability of accessibility metrics across 25 repeated crawls.}\label{tab:stability}
\begin{threeparttable}
\begin{tabularx}{\textwidth}{l XXX XXX XXX} \toprule
  & \multicolumn{3}{c}{\textbf{Pre-interaction}} & \multicolumn{3}{c}{\textbf{Post-accept}} & \multicolumn{3}{c}{\textbf{Post-reject}} \\
  \cmidrule(lr){2-4}\cmidrule(lr){5-7}\cmidrule(lr){8-10}
  \textbf{Metric} & \textbf{Range} & \textbf{Mean} & \textbf{\gls{cv}} & \textbf{Range} & \textbf{Mean} & \textbf{\gls{cv}} & \textbf{Range} & \textbf{Mean} & \textbf{\gls{cv}} \\ \midrule
  Lighthouse score & 100 & 100.0 & $0.0$ & 98 & 98.0 & $0.0$ & 98 & 98.0 & $0.0$ \\
  WAVE errors & 1--3 & 2.8 & $19.5$ & 1--3 & 2.8 & $19.5$ & 1--3 & 2.8 & $24.0$ \\
  WAVE contrast errors & 0 & 0.0 & $0.0$ & 0 & 0.0 & $0.0$ & 0 & 0.0 & $0.0$ \\
  WAVE alerts & 1 & 1.0 & $0.0$ & 1.0 & 1 & $0.0$ & 1 & 1.0 & $0.0$ \\
  WAVE features & 12--13 & 12.7 & $3.6$ & 12--13 & 12.7 & $3.6$ & 12--13 & 12.6 & $4.0$ \\
  WAVE structural elements & 9--13 & 12.7 & $8.7$ & 9--13 & 12.7 & $8.7$ & 9--13 & 12.5 & $10.6$ \\
  WAVE ARIA & 450--490 & 481.4 & $2.1$ & 430--473 & 464.6 & $2.2$ & 430--473.0 & 463.8 & $2.7$ \\
\bottomrule\end{tabularx}
\begin{tablenotes}\footnotesize
  \item Range shows min--max; \gls{cv} is the coefficient of variation (\%).
\end{tablenotes}
\end{threeparttable}
\end{table}

\begin{figure}[H]
  \centering
  \includegraphics[width=0.7\linewidth,keepaspectratio]{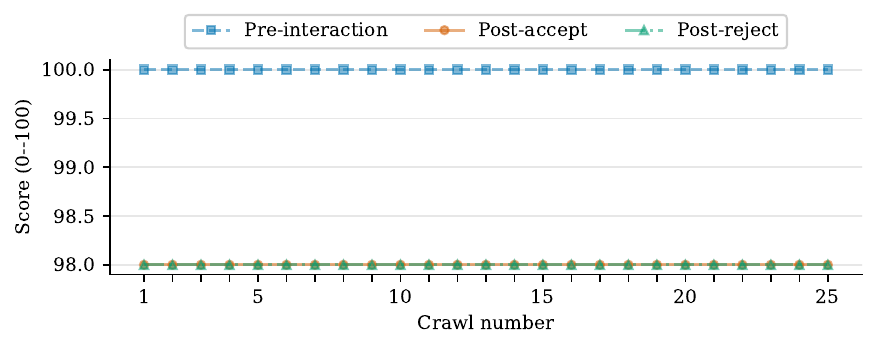}
  \caption{Lighthouse accessibility score across 25 repeated crawls of \protect\url{google.com}.}
  \label{fig:stability_lh}
\end{figure}

\begin{figure}[H]
  \centering
  \includegraphics[width=0.7\linewidth,keepaspectratio]{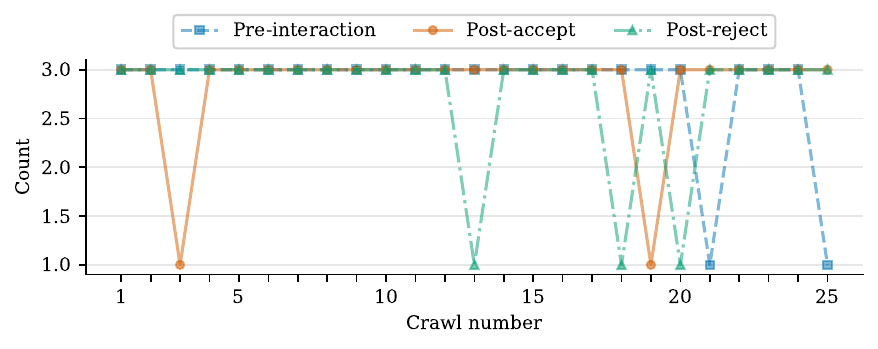}
  \caption{\gls{wave} error count across 25 repeated crawls of \protect\url{google.com}.}
  \label{fig:stability_wave_error}
\end{figure}

\begin{figure}[H]
  \centering
  \includegraphics[width=0.7\linewidth,keepaspectratio]{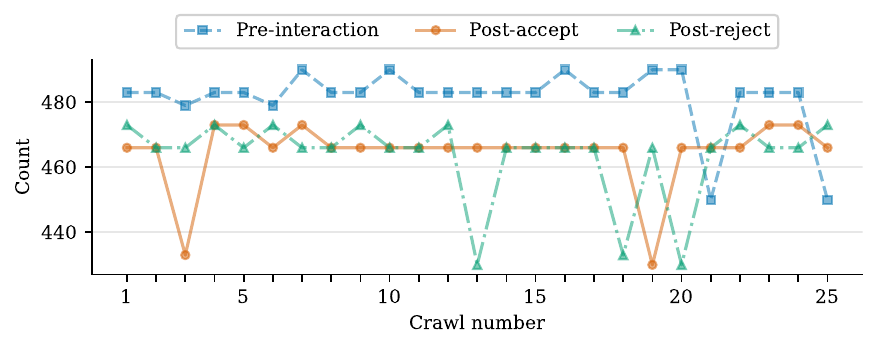}
  \caption{\gls{wave} \gls{aria} element count across 25 repeated crawls of \protect\url{google.com}.}
  \label{fig:stability_wave_aria}
\end{figure}

\subsection{Accessibility tests}
\autoref{tab:stability} summarises the min, max, mean, and \gls{cv} for each metric across all three interaction phases.
The Lighthouse accessibility score was perfectly consistent across all 25 crawls in every phase (pre-interaction: 100, post-accept: 98, post-reject: 98, see \autoref{fig:stability_lh}), as were \gls{wave} contrast errors and \gls{wave} alerts, all of which showed zero variance.
The remaining metrics showed only minor variation, with a maximum \gls{cv} of 24.0\,\% (\gls{wave} errors, post-reject), as shown in \autoref{fig:stability_wave_error}. This level of variation is attributable to dynamic JavaScript content on the page rather than measurement instability: Google's search page renders components asynchronously, causing small fluctuations in the number of \gls{aria} elements detected between crawls.
Notably, the mean \gls{wave} \gls{aria} count decreases from 481 pre-interaction to 465 post-accept (\autoref{fig:stability_wave_aria}), consistent with the cookie notice's \gls{aria}-annotated elements being removed from the \gls{dom} upon dismissal.
Overall, these results demonstrate that our crawler produces highly reproducible measurements.

The Lighthouse score drop from 100 to 98 is explained by a single, consistently reproducible audit failure: \emph{Document does not have a main landmark.} (\texttt{landmark-one-main}). This audit failed in 25/25 post-accept crawls and 25/25 post-reject crawls, but was never flagged pre-interaction. The cookie notice serves as the page's sole \texttt{<main>} landmark; its removal upon dismissal leaves the document without a main landmark, triggering the audit. This finding was perfectly consistent across all crawls, with no exceptions.

Screen reader metrics were evaluated across all 25 crawls.
All of the following criteria passed in every crawl without exception: (i) \emph{Readable}, (ii) \emph{Immediately Read}, (iii) \emph{Keyboard Navigable}, (iv) \emph{Link or Button Purpose}, (v) \emph{Abbreviations Explained}, (vi) \emph{Page Titled}, (vii) \emph{Cookie Notice Titled}.
For criterion~(ii) \emph{Immediately Read}, the first cookie-related keyword appeared at exactly 10 words into the \gls{nvda} transcript in all 25 crawls (min=10, max=10), with all 25 applicable crawls passing the 30-word threshold. The perfect consistency of this metric across crawls further confirms the stability of the screen reader analysis.
\end{appendix}

\end{document}
\endinput